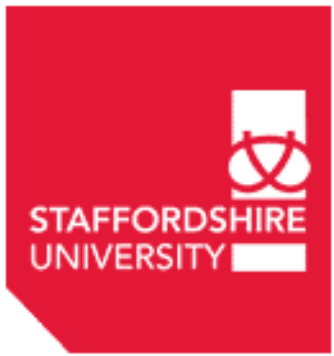
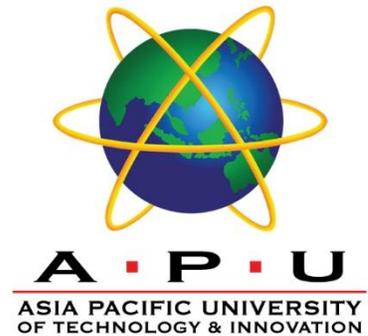

# Synthesizing Cough Audio with GAN for COVID-19 Detection

## FYP Report

By **Yahya Saleh**

B.Sc. (Hons) in Intelligent Systems

May 2023

# Table of content

# List of Figures

# Table of Tables

# 1   Introduction

## 1.1   Project background

Health care is one of the most impactful and important fields in today's industry, it is essential to people's lives and wellness. Yet once again, during the Covid-19 pandemic, the shortcomings of the industry were highlighted as the outdated system governing health care struggled to keep up with the ever-growing demand for medical services.

For the field of artificial intelligence and specifically machine learning and deep learning, many promising findings that are revolutionizing different sectors of life as Transportation (Khan, 2022; Mokayed, 2021; NG, 2015), document and financial analysis (Nikolaidou, 2022;  Kanchi, 2022, Mokayed, 2014), biometric verification (Marzuki, 2011), natural language (Alkhaled, 2023; T Adewumi, 2022), smart city and security (Mokayed 2022;), industrial application (Masri, 2021), drones and robotics (Mokayed, 2021) and health care (Voon, 2022). Many of those promising findings in different sectors and mainly in health care are remained theoretical or have fallen short due to the scarcity of medical datasets (S, 2021). Recording medical data is difficult and expensive and thus data is scarce within the medical field which made deep learning less effective as by its nature it requires a huge quantity of data to learn the underlying latent features correctly and show acceptable performance. To solve this problem, some researchers started turning to deep generative models for image synthesis, where the goal is to build a generative adversarial network, or the like, to generate new synthesized medical data that fall within the domain of the dataset effectively increasing the size of the dataset.

Likewise, for this final year project, the goal is to add to the published works within data synthesis for health care. The end product of this project is a trained model that generates synthesized images that can be used to expand a medical dataset (Pierre, 2021). The chosen domain for this project is the Covid-19 cough recording which is have been proven to be a viable data source for detecting Covid. This is an under-explored domain despite its huge importance because of the limited dataset available for the task. Once this model is developed its impact will be illustrated by training state-of-the-art models with and without the expanded dataset and measuring the difference in performance. Lastly, everything will be put together by embedding the model within a web application to illustrate its power. To achieve the said goals, an extensive literature review will be conducted into the recent innovations for image synthesis using generative models.

## 1.2   Problem Context

- **Scarcity of medical data**: Medical data is scarce and this is stifling progress within the field (Maier, 2020).

- **Medical data are costly**: with no easy way to obtain medical data acquiring them is costly (Pierre, 2021).
- **Discrete Medical Data**: Another concern within health care is protecting the privacy of their patients' information from adversaries. While there are ways used to hide the information of the patient whose medical data belong, those algorithms can be exploited. With GANs Discrete medical data can be created such that it cannot be tracked back to the patient (Pierre, 2021).
- **Lack of experts**: Another usefulness that image synthesis provides is that it makes up for the limited number of experts to annotate the data once it is generated which helps train models for supervised learning tasks (Yi et al., 2019).

*1.3    Rationale*

Having understood the background of the project as well as its problem context the rationale behind the proposed solution becomes vividly clear. Health care is one of the most vital domains that technology should improve and revolutionize, yet AI-based solutions have fallen short within the domain due to the scarcity of datasets within the said domain. Acquiring data within the field of health care is difficult because it is costly and prone to many privacy complications. Looking at the Covid-19 domain more specifically not only is it subject to the same limitation that stifles the production of AI-based solutions, but it is also a new virus that is not completed comprehended, and thus even less data is available.

*1.4    Potential benefits*

The critical analysis of image synthesis literature performed in the initial phases of the project will give researchers better insights into the current breakthroughs in image synthesis. Furthermore, the Developed GAN model which will be used to synthesize Covid-19 cough audio will provide valuable insights by examining the latent features the model learned to synthesize the said data. Additionally, using the synthesized data to improve the performance of covid classification through coughing will improve the covid screening process. Overall, the benefits of this project are very promising to fields of Artificial intelligence and health care.

*1.5    Targeted audience*

The targeted audience for this project would then be artificial intelligence developers and biomedical engineers that build AI-based solutions for health care, as well as AI experts and developers who can use the proposed improvements to reach new findings and build innovations. Furthermore, the synthesized data can provide insight into the characteristics of Covid-19 and help experts understand the virus and its impact better.

## *1.6    Scope and objectives*

### 1.6.1    Aim

To Build a scalable Generative Adversarial Network for Image synthesis to provide a means for generating synthetic medical data within the domain of Covid-19 cough audio to further the progress of AI within health care and inspire innovations across all domains.

### 1.6.2    Objectives

The aim of this project can be broken down into the following individual objectives, each of which has its own set of sub-tasks to be fulfilled.

- **Analyzing the existing literature in GANs and Covid detection**

A literature review of the recent innovations within the field of image synthesis needs to be done to understand the latest algorithms used for the tasks to guarantee an improved GAN structure for the selected task. Following that is a literature review of the latest models and AI-solution deployed to aid in the Covid-19 pandemic. Lastly, understanding how experts analyze cough audio for covid and the features that are present in such an audio piece.

- **Implementing the Generative Adversarial network**

Exploring the power of GANs and their ability to generate synthetic data. Then building a GAN that synthesis new images on a general dataset. This generalized approach will help in perfecting the GAN's performance without worrying about the impact of the dataset or domain on the performance which gives a reliable measurement of the wellness of the model. The built GAN structure will be used to synthesize images within the selected domain. After viewing the results of the GAN the next step is Debiasing the model for a more representative output.

- **Illustrating the impact of the synthesized data**

The impact of the model will be illustrated by improving the performance of state-of-the-art models and evaluating the degree of improvement achieved with the expanded dataset.

- **Embed the finalized model into a web application to illustrate the power of the proposed neural network.**

Once the work has been compiled and documented, its impact and power will be illustrated via a web application.

## 1.7   Deliverables

The following are the expected deliverables of the project:

- To Analyze the existing literature in GANs and Covid detection
- To Implement a Generative Adversarial network for cough data synthesis that captures the underlying latent features of the covid present in the audio.
- To improve the performance of models on the task of covid detection through cough
- To develop a web application that leverages the proposed pipeline

## 1.8   Nature of challenges

The biggest challenge for this project and generative adversarial networks, in general, is ensuring that the discriminator does not overfit the data and that the generated new data is distinct.

The main limitation of the project is that the generated model would be created to fit a specific dataset, which given the difficulty of the task within the domain of health care does not guarantee that the model's architecture will be scalable. Nevertheless, the structure of the model will most likely inspire new architectures that are scalable or are configured to fit a new domain within health care.

The main learning objective of this project is understanding how deep generative models generate latent features, which are then used to construct slightly different images that fall within the same domain. Understanding how this process works and the stochasticity brought by introducing distributions to the model instead of assuming the parameters is the key to building the GAN.

## 1.9   Overview of the investigation report

The purpose of this investigation report is to perform the preliminary research required to start the project. Through its chapters, this report will document the technical and theoretical aspects that must be understood to accurately assess the feasibility of the project and its impacts.

This report is divided into 7 chapters. The first of which is the introduction chapter which documents the motivation behind the project, the potential gains from the project, and the goals of the project. The second chapter provides a detailed analysis of literature where the author discusses the various innovation with both the selected domain of interest and using the algorithm proposed.

For this report, the domain will be the Covid-19 cough audio recordings and the various innovation reviewed will be within the field of image synthesis for health care.

The third chapter concerns technical research where the chosen tools are documented. The fourth chapter documents the selected methodology for the project. The fifth and sixth chapters are concerned with requirement validation which for this project would be assessing the accuracy of the synthesized images through the help of domain experts. Lastly, the final section summarizes the documented finds and reflects on the proposed idea behind the project: its feasibility and impacts. The conclusion of the is report is to determine whether this project is worth the endeavor that is yet to come.

## 1.10  Project plan

The first step to getting started with the project is to acquire the domain and technical knowledge. The $17^{th}$ of January, 2022 to the $30^{th}$ of the same month will be allocated to Domain selection and understanding and dataset selection and documentation. Then from the $6^{th}$ of February to the $20^{th}$, extensive research on GANs and the recent innovations of GAN on dataset expansion will be carried out.

With both domain and technical knowledge acquired the next step is to start programming and building the model's architecture. To decouple the process and focus on building each block separately, the first task would be to implement a GAN for dataset expansion from scratch and apply it to a generic dataset, this way the sole focus is on making the model work without worrying about the impact of the domain and the nature of the dataset. Of course, the effectiveness of the technique will be tested as stated previously. This whole task will be carried out from the middle of February till the middle of March.

With a working GAN architecture, the next step is to implement the model on the domain dataset and measure the performance. Since a working GAN has been implemented, one might think that this task will be simple, but it is the hardest because the model needs to be debiased, and synthesizing data within the health care domain is much other than a generic domain. This is task will take a month from mid-March to mid-April. With a debiased functioning model, the next step is to measure the impact on the selected domain and document the results. Since various experimentations and testing will take place, this task will take another month lasting till mid-May.

Once a paper on the impact of GANs for dataset expansion within health care is ready based on the illustrated work, the next step is to improve upon the results by proposing improvements to GAN architecture and documenting the findings after rigorous testing. This task will last a month till mid-June.

Finally, once the work is documented and ready for publishing, everything functionality will be illustrated by embedding the finalized model within a web application, a task that will start from mid-June and last till the first week of July.

Those illustrated sets of tasks follow the management structure of a DevOps deployment methodology where the tasks are decoupled and each is completed and integrated independently.

## 2 Literature Review

### 2.1 *Generative adversarial networks (GANs)*

Generative adversarial networks as first defined by Goodfellow et al. (2014) are a set of two neural networks used for generative modeling where the network draws samples from the data distribution in an implicit way. Drawing samples implicitly means that rather than mapping the underlying distribution explicitly, the model is taught to create samples that match the distribution of the data. This is achieved through the use of two neural networks the generator, $G$, and the discriminator, $D$.

As shown in the figure above the generator neural network takes in as input random noise, $z$, that is sampled from a prior distribution, commonly a Gaussian distribution denoted as $p(z)$, and outputs data, $x_g$, that is similar to the real sample, $x_r$, which is drawn from the real data distribution, $p_r(x)$ (Yi et al., 2019). For the case of image data, for example, the generator would take in a random noise and output an image that has visual similarities to images in the dataset.

*Figure 2.1: Vanilla GAN strcture (van de Wolfshaar, 2018)*

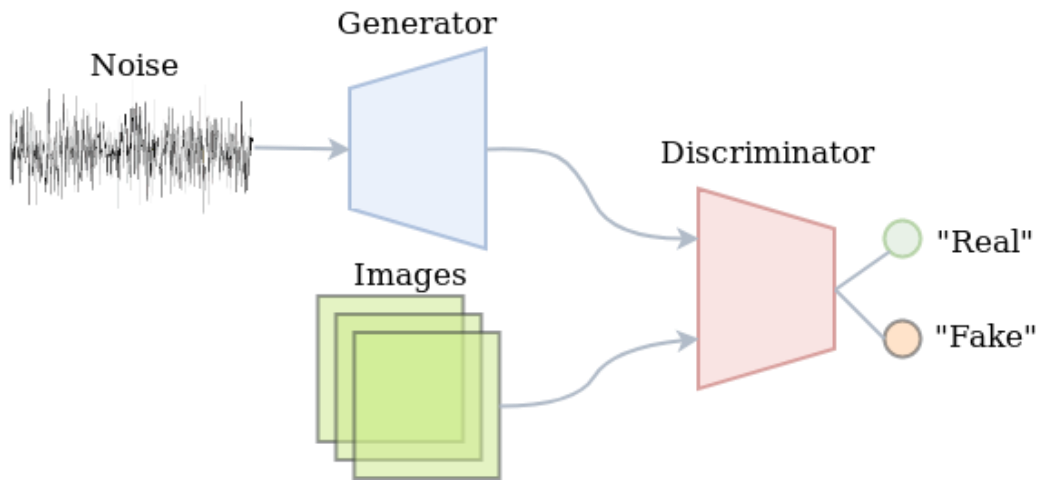

The discriminator on the other hand takes in data from the selected domain, such as images. This data could either be generated by the generator or from the dataset itself. And the task of the discriminator is to determine whether the given image is real, i.e. belongs to the dataset, or fake. Thus the output of the discriminator is a single value representing the probability that the input is real or fake.

The two models are adversarial where the generator tries to fool the discriminator and the discriminator tries to detect data generated by the generator. What makes this structure unique and powerful is that the discriminator teaches the generator how to better improve its generated quality as the gradient information is backpropagated from the discriminator to the generator (Yi et al., 2019). The following functions express the training objectives of the two models:

*Equation 1: Training objects of the generator and discriminator models*

$$\mathcal{L}_D^{GAN} = \max_D \mathbb{E}_{x_r \sim p_r(x)}[\log D(x_r)] + \mathbb{E}_{x_g \sim p_g(x)}[\log(1 - D(x_g))]$$

$$\mathcal{L}_G^{GAN} = \min_G \mathbb{E}_{x_g \sim p_g(x)}[\log(1 - D(x_g))]$$

The equation shows that the discriminator maximizes the probability returned when given data from the dataset, $\log D(x_r)$, and minimize the returned probability when given a generated image which is represented as maximizing the difference between one and probability returned given a generated image, $\log(1 - D(x_g))$. This makes the discriminator a binary classifier with a maximum likelihood objective. The generator on the other hand minimizes the probability that the data it generated is fake, $\log(1 - D(x_g))$.

Goodfellow et al. (2014) states that if the discriminator is trained optimally before the generator updates its weights, then minimizing the generator's loss function, $\mathcal{L}_G^{GAN}$, will minimize the Jensen-Shannon divergence between the generated data distribution, $p_g(x)$, and the real sample distribution, $p_r(x)$.

While there are many challenges in training the GAN model optimally as stated by Yadav et al. (2018), the power and potential of the algorithm made it the center of image synthesis. Many research introduced a variation on GAN architecture to make it more robust and expand its potential. For example, different loss functions for the discriminator were introduced to stabilize the training process since handling the discriminator is one of the biggest challenges when training a GAN. Most notably is the Wasserstein distance as it is the most popular (Arjovsky et al., 2017; Gulrajani et al., 2017).

Other authors applied variations to the generator to enable the GAN to perform new tasks such as image translation. Such capability is achieved by giving the generator extra information, $c$, to drive it into creating output, $x_g = G(z, c)$, with a set of desired properties. Such GAN is referred to as a conditional GAN (Mirza & Osindero, 2014). One popular application of this alteration is the aforementioned image-to-image translation where the first framework was proposed by Isola et al. (2016). The following figure showcases some examples of image-to-image translation from Zhu et al. (2017) the paper that proposed the architecture of cycle GAN a very popular and power variation:

*Figure 2.2: Image-to-image translation (Zhu el al., 2017). given two domains the GAn leans to translate an image from one domain to the other.*

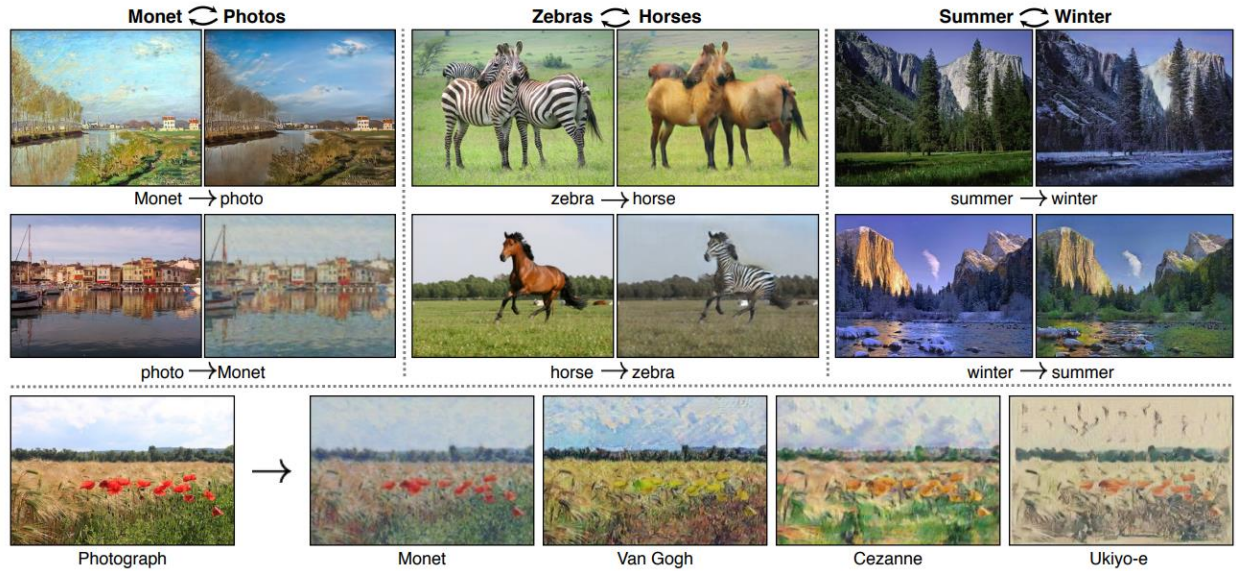

## 2.2 Covid-19 detection through coughs

To the best of our knowledge, there has been no paper published on synthesizing cough audio for covid-19 detection; however, many published works show the need for such research. In the wake of the covid-19 pandemic, many researchers explored the role machine learning can play in providing efficient screening. Many of those researchers introduced machine learning-based solutions for detecting the virus through a person's cough.

The work by Darici et al.(2022) tested multiple models for the task of detecting covid given cough audio. The study found a possibility for bias given the limited dataset and concluded that more data is required to improve the performance of the model. Another work by Lella & PJA (2021) applies a one-dimensional CNN for the same problem. However, it is worth noting that the authors have collected around five thousand samples for the research with only 300 of those samples declared as Covid-19 positive patients.

Looking at those findings it is clear that there is a huge scarcity of data within this domain, and many entities took action to mitigate this limitation. The DiCOVA challenge, its first launch (Muguli et al., 2021) and second (Sharma et al., 2021), was designed to encourage researchers to analyze recordings of people infected and not infected with covid. However, much like the previous datasets, the ones featured in the tracks were fairly biased with the first track dataset having 965 samples only 172 of which are Covid-19 positive, and the second track having 471 only 71 of which are covid positive (Sharma et al., 2021). And indeed this bias was reflected in the performance of teams in the challenge as they stated in their paper (Sharma et al., 2022) that the top 4 teams fell short of the $\geq 70\%$ (at a specificity of 95%) stated by the World Health Organization. Although a simple combination of the scores from models developed by those teams meets the stated goal the authors state "there is scope for further development in future" (Sharma et al., 2021).

The efforts in proving suitable datasets for covid screening through audio to the community continued, most notably the work by Orlandic et al. (2021) who proposed the COUGHVID crowdsourcing dataset a large-scale dataset for cough analysis. This dataset consists of 20,000 samples with 1,010 being samples of people "claiming to have COVID-19" (Orlandic et al., 2021). The increased dataset size will certainly improve the performance of the models; however, the bias against covid-19 patients still resides. And it turns out that the first explored work by Darici et al.(2022) used the COUGHVID along with the Coswara dataset (Sharma et al., 2020) and still reported bias in the model's performance.

# 3    Technical Research

## 3.1    *Programming language*

Various programming languages can be used for deep learning and picking the right programming language for the task is crucial to ensure smooth development and progress. A good way to find out the best programming language for a field is to get the opinions of experts within that field. When it comes to machine learning and deep neural networks **Python** takes the lead with 57% of all data scientists and machine learning experts using it for development, and 33% prioritizing it (Nation, 2019). The reason for this popularity is the various advantages python provides for machine learning developers, including a rich library ecosystem, platform independence, and a huge community and support (Great Learning Team, 2021).

Another option that many developers consider for machine learning and deep learning is **R**; however, as it turns out R is not very popular in the field, with just 31% of developers using it and only 5% prioritizing it (Nation, 2019). While this may come as a shock for some developers, those results are very consistent with the purpose of the language. R is designed for data analysts and scientists with mathematical backgrounds rather than ones with programming or computer science backgrounds (r-project, n.d.). In that light, it makes sense that R will not be tailored or well suited for deep learning or the like because that falls outside the range of its purpose.

A more suitable choice for developers is **C/C++** but it remains a distant second to Python. While C++ might seem like a more appealing option for computationally demanding tasks such as deep learning, Python is easier to learn and easy to use in practice (Rohner & Netguru, 2021). Furthermore, most published papers and works in the field of machine learning use Python as the language of choice to implement their models (Rohner & Netguru, 2021) which gives Python an edge.

As such the chosen programming language for this project is Python, a very popular programming language that has huge support for deep learning and a huge community base for assistance. Using Python will allow the developer to implement and improve on other implemented work easily as in all likelihood the models implemented in the literature were developed using Python.

## 3.2    *Libraries*

As stated in the previous subsection, Python has a rich library ecosystem that enables the developer to leverage prewritten code for each stage of the development. The main library that will be used in the project is Tensorflow, a powerful library for generative deep learning that uses the Keras library as an API endpoint for various machine learning tasks.

Other libraries might be used as well for the various stages of development, for example, OpenCV can be used when preprocessing the data and then the preprocessed data can be stored in a pandas frame for analysis. Furthermore, NumPy arrays provided in the NumPy library are known to be more efficient and faster in their performance compared to the list data type built into Python.

### 3.3 Integrated Development Environment

Rather than choosing an Integrated Development Environment for this project, visual studio code which is a code editor will be used instead. Vscode is the most popular code editor as shown in the 2021 Stack overflow Developer survey (Stack Overflow, 2021), shown below, where 71% of developers use it as their main code editor of choice.

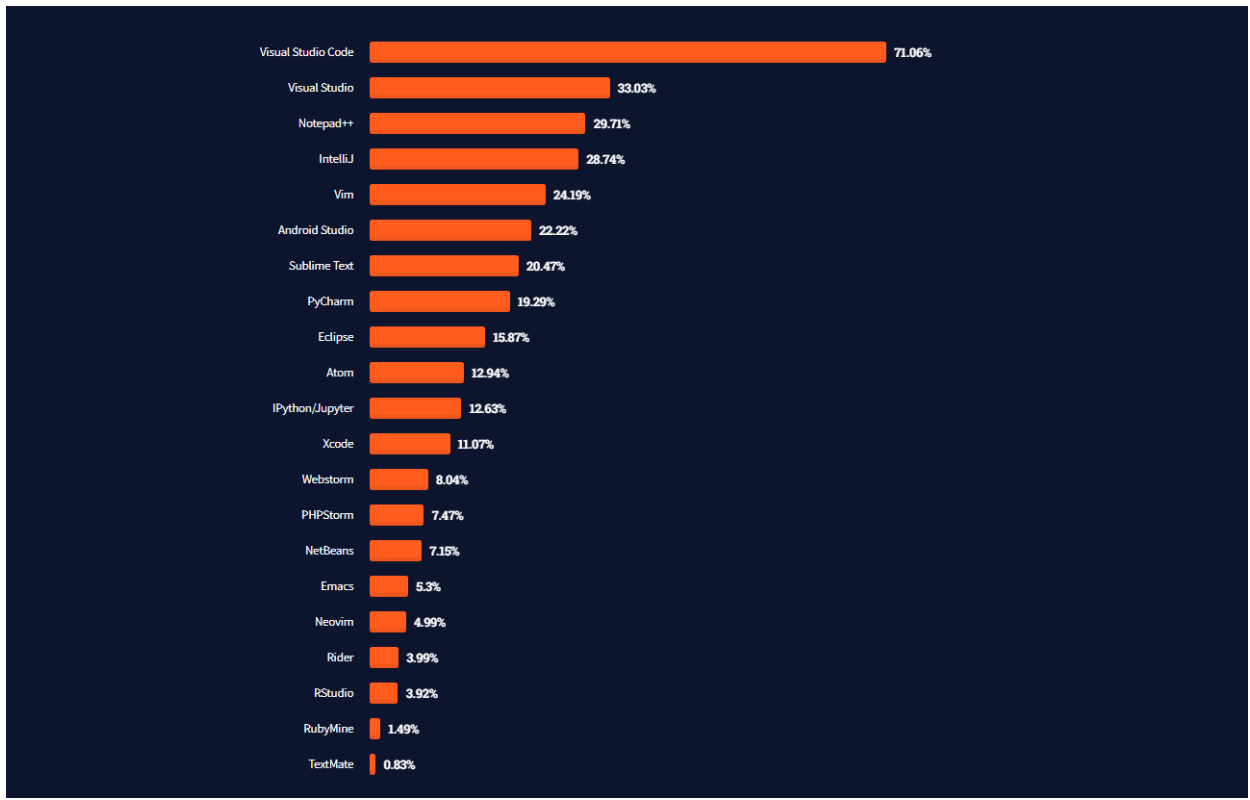

*Figure 3.1: Stack overflow Developer survey IDE of choice across all developers (Stack Overflow, 2021)*

vscode is so popular among developers for its lite and flexible build that makes everything within one editor including version control. Vscode has many community-developed and supported extensions that make development very efficient with snippets, code highlights, and IntelliSense being used. Python and Python supporting extensions are the most installed extensions which show the popularity of Python within vscode. Furthermore, vscode has the best support for web development which makes it the perfect code editor for completing the project and then developing a web application to showcase the work.

## 3.4   Algorithm chosen

As discussed throughout this investigation report the chosen algorithm for image synthesis is **Generative adversarial networks (GANs)**. GANs generate very accurate and detailed synthesized images that convincingly look like they fall within the selected domain using a novel algorithm with two models first introduced by Ian Goodfellow (Goodfellow et al., 2014). GAN algorithms learn irregularities or patterns from the inputted image or video such that they can generate new images or videos that look like they were in the dataset (Brownlee, 2019). The following figure, for example, uses a GAN model to generate artificial faces that did not exist before using two inputted images: denoted as the source and destination:

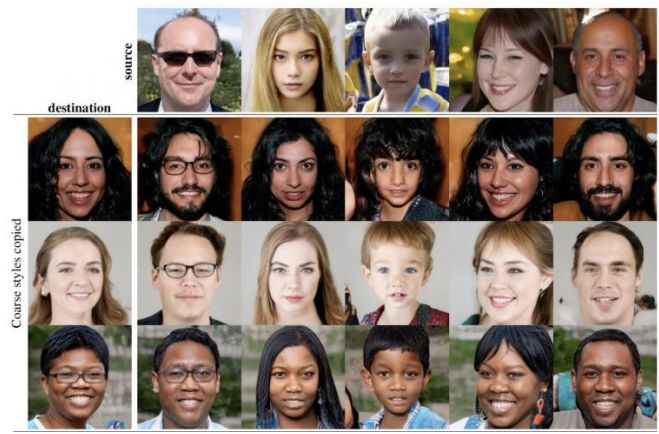

*Figure 3.2: Generating Realistic Artificial Faces (Horev, 2019)*

GANs work by training two deep neural networks: a **generative model** and a **discriminative model**. The former is trained to generate new examples using the inputted data, and the latter is trained to discriminate and determine if a given data is from the data set or not. The two models get trained simultaneously until the discriminative mistakenly believes that at least half of the data generated by the generative model is real (Brownlee, 2019).

## 3.5   Hardware

The hardware requirements for this project have been tailored to meet the required hardware specs for deep learning, as deep learning is very computationally demanding and is the core of the project. The first and most important factor to consider is the general processing unit (GPU), they are more popular and used more often for deep learning since they have a better capacity of handling simultaneous computation than CPUs. Here are the requirements for a good GPU suited for deep learning:

- memory bandwidth
- number of cores
- processing power
- Video RAM size

A minimum of 8 GB GPU is ideal for deep learning, and the ideal GPUs to pick from are 3080/3080ti/3090 which all belong to NVIDIA, whose recent processing cards come with AI-enabled functionalities and support popular deep learning frameworks (Ben-Zvi, 2022). The second factor to consider is RAM, the higher the RAM the greater the number of data that the system can handle (Zindi, 2020). Usually, training a deep learning model can be carried out with 8 GB of RAM, but a 16 GB RAM or higher is much recommended. The last factor is the CPU. The minimum CPU that guarantees constant performance is the Intel Core i7 processor. Anything of equivalent processing power or higher should be sufficient (Zindi, 2020).

Those requirements are demanding and are not available on any normal computer. There are a couple of options to meet those requirements: first, using google's hardware through google Colab notebooks. Second, renting a workspace from amazon that provides a virtual machine to be accessed through any device where the virtual machine has the required hardware specifications.

## 3.6   Database

There is no need for an integrated database for this project; however, if the web application requires storing data for future use, fire cloud will be the database of choice as a free very flexible cloud storage that uses a NoSQL structure for data storage.

## 3.7   Web framework

The chosen web framework for the backend of the web application that will demonstrate the work accomplished in this project is Flask. Flask is a Python-based web development framework that is much like Django but is popular for simple projects. The reason is it requires far fewer configurations than other frameworks and does not worry about the web project having various apps which makes it easy and simple to use.

Since the Flask framework is Python-based it is will perfect for rendering this project that will be developed in Python, where the trained models can be easily integrated without any need for model handling code in another language.

Lastly, the frontend will be developed using vanilla Javascript if any programming logic is required on the frontend of the web application.

# 4    Development Methodology

As stated throughout this investigation report the main goal of this project is to contribute to the domain of generative modeling and health care by proposing an improvement on the generative adversarial network architecture that will synthesize medical data, expand medical datasets, and further the development and integration of AI in the health care industry. As such this project requires a unique development Methodology that is more centered around developing and optimizing the generative model and measuring its impact.

## 4.1    Proposed Development methodologies for deep learning

There have been different research and proposed development methodologies centered around neural network development and its integration into the software. A research paper titled "A Methodology to Develop AI Software in an Organization" (Rago, 2018) discussed the importance of adopting system development methodologies to accommodate for the integration and use of AI in the system as an embedded component. The proposed methodology included the known phases of a system development life cycle where the stages in order where:

1.  Plan – "The purpose of the initial activities is to analyze corporate culture and conduct a high-level process analysis" (Rago, 2018);
    a.  Competences management
2.  Design – "you go to the macro-level definition tasks of the process model. In this way it is possible to carry out the gap analysis, which determines the guidelines for subsequent activities." (Rago, 2018);
3.  Build – "The Build phase involves the realization of software and the development of knowledge starting from the big data of the company.";
4.  Operate -  the implementation of the complete system.

This methodology is more centered around software development itself and does not treat the integration of neural networks as the main component. Looking at the proposed stages it is clear that the neural network aspect is treated as a secondary component one that can be done as part of the system's life cycle.

Another research paper titled "Software Engineering Methods for Neural Networks" (Senyard et al., 2004) introduced methods that can help AI developers deviate from the ad-hoc method, i.e. trial and error, and build models in a methodological way that makes the results repeatable and constant. The authors of this study define the common approach for developing the neural network as comprising of the following stages:

1.  Choosing a problem representation and fetching an associated dataset;
2.  Setting the performance expectation of the model;
3.  Training the neural network until the expected performance is met;
4.  Updating the model's parameters to improve the model.

They then went on to introduce some order to the development process by proposing that rather than defining a single problem representation along with a single dataset various problem representations must be identified depending on the developer's judgment. Then each problem representation along with their respective dataset is ranked by reliability and the model is trained on each until the goal is met. For deep learning experts, the description of the existing development process and the proposed process by this paper may seem off and incorrect.

### 4.2  Shortcomings of development methodologies for deep learning

Most research papers, including the ones discussed, miss the mark when it comes to developing neural networks. This could be because the developers introducing the development methodologies come from a software engineering background and thus lack the detailed knowledge and experience behind developing neural networks, which is ever-changing and evolving, and instead they try to frame the problem that a neural network will solve as a system development problem which at best is not a valid approach.

Neural network development is not a component interrelated to the system that can be developed as part of the system's life cycle as implied by the "A Methodology to Develop AI Software in an Organization" (Rago, 2018) paper. Instead, it is an independent component that requires a great deal of iteration and testing to deploy a reliable model. The trained and complete model is what then becomes a component that can be integrated as part of the system or for commercial use.

Another reason that is hard to find methodologies that perfectly define the development process of neural networks, is that neural networks or deep learning more specifically is a huge field with various models and architectures used for distinct applications such as image classification, resolution improvement, autonomous cars, image synthesis, reference learning, etc. And those various applications deal with various data types and records such as regular numeric data, image data, sequential data, and generative data.

However, there still exists a need for an organized development methodology to improve the pace of development and provide a road map for deep learning. Many machine learning experts will highlight that issue and stress the need for a structured methodology for developing any AI model (Pinhasi, 2021); however, since they have experience with AI development they will point out how even the most agile of methodologies still leave something to be desired when used for machine learning development. The following is a summary of some of the popular development methodologies and why they are not completely suitable for machine learning let alone deep learning:

| Methodology | Conclusion | Reason |
|---|---|---|
| Agile | does not work as-is | Agile practices were developed for engineering and don't generalize as well for AI (Pinhasi, 2021) |

| Scientific Research Methodology | Not a complete solution | Offers useful techniques to deal with uncertainty, but the methodology was not designed for commercial release (Pinhasi, 2021) |
|---|---|---|

*Figure 4.1: Shortcoming in popular methodologies*

## 4.3    Chosen methodology for deep learning

As illustrated machine learning problems in general often fall between research and deployment as they require a great deal of understanding to pick the right algorithm and improve upon. At the same time, the development must be robust and reliable as AI is used for critical problems. And since deep learning is used to solve a wide array of problems, there is no concrete development methodology to be applied. Instead, experts rely on their judgment to develop a plan that will in all likelihood adopt as the results and findings of the model unfolds.

Nevertheless, machine learning experts still try to come up with a generalized workflow for machine learning as shown in the flow chart below. However, those workflows will probably not be applicable for deep learning where neural networks are used to deploy various techniques that make the stages and problem structure completely distinct.

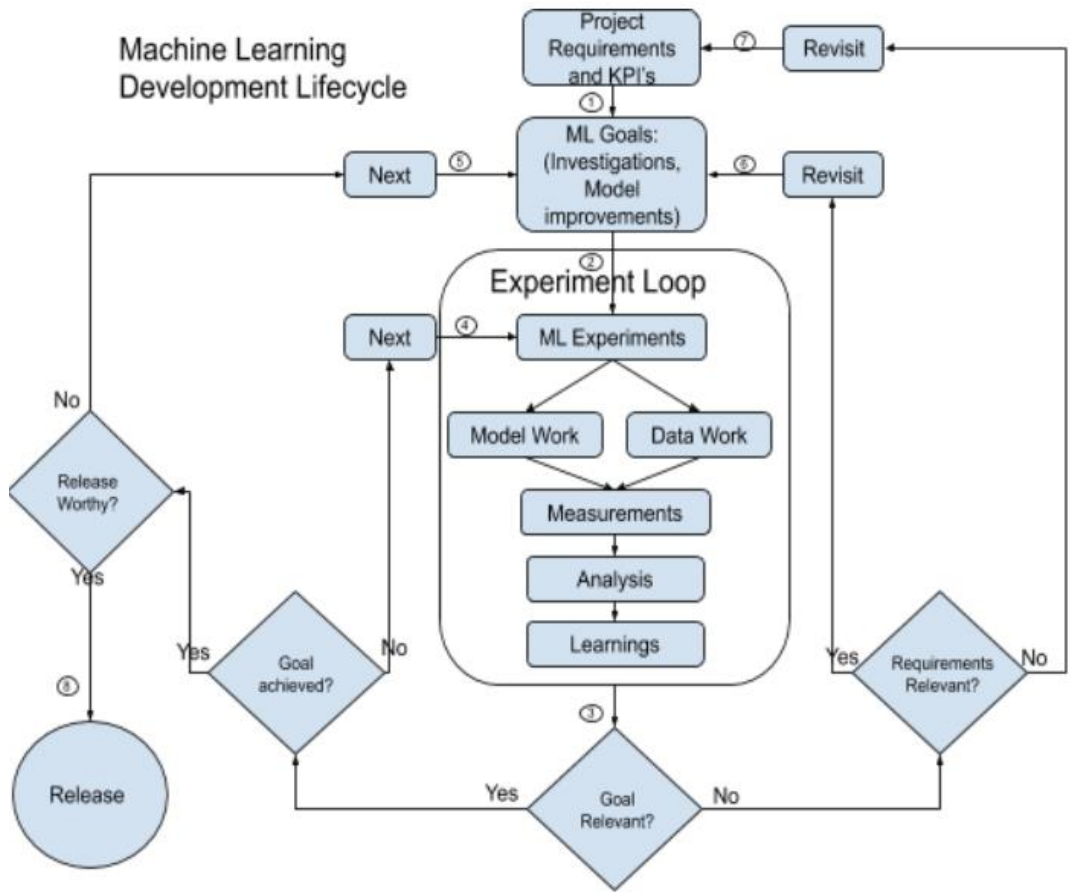

*Figure 4.2: Draft Machine learning development life cycle (Pinhasi, 2021)*

As such the development methodology for this system will be a custom methodology following a simple set of stages to those of Machine learning development. The stages are as illustrated in the project's plan.

- **Domain and dataset selection**

This stage deals with selecting the best domain of interest within the field of health care, which at the time is Covid-19 chest Xrays, and selecting the appropriate data to be synthesized.

- **Dataset preprocessing and analysis**

This stage is concerned with loading the data preprocessing it and understanding the latent features of the data which will help the developer when evaluating the quality of synthetic data generated by the trained GAN.

- **Literature review**

This stage is where critical analysis is performed to understand the recent innovations in generative adversarial networks and how these innovations and improvements are used within the field of healthcare. This critical review will provide valuable insight for developing the generative adversarial network and understanding the shortcomings in the specific field.

- **Building a general GAN model**

This stage is concerned with understanding the algorithm behind the GAN model to better improve the proposed improvement and find the best alteration to the model to suit the task at hand.

- **Building a model to synthesize image data**

This is the process where everything is put together as the model is and trained to produce synthetic data for the specified domain.

- **Model evaluation**

In this stage, the model's performance is critically evaluated as the developer ensures that the right features have been synthesized and that the generated data can be useful. This will be the most rigorous process as various improvements will be applied to the model and experts will be asked to review the results.

- **Model use for dataset expansion**

Then the model will be used to expand a given dataset and the performance of the model will be measured comparatively with and without the expansion on state-of-the-art models that have the best ability for identifying and leveraging the latent features of the data.

- **Model deployment**

Once the model is complete the last set is to deploy the model for commercial use and build the aforementioned web application to illustrate the model's ability and power.

## 5   Reflection

Overall Synthesizing Cough Audio with GAN for COVID-19 Detection is very promising research that has yet to be explored. The potential behind this idea is huge and as such requires a great deal of work, but the outcome is guaranteed to be outstanding.

At the end of the investigation stage of this project, the idea and research topic was finalized, an appropriate dataset was selected, and adequate research on the technical and domain aspect of the idea have been conducted. There is still room for change and improvement in the plan as the developer explores the idea and the challenges behind it, but overall the plan is set and will be moved forward.

# 6    Dataset

The selected dataset for cough augmentation is the coughvid dataset (Orlandic et al., 2021) the large Covid-19 audio dataset. The coughvid dataset consists of 25,000 cough samples of different patients with diverse age, gender, and geographical locations, 1155 of which belong to a covid-19 positive patient. Furthermore, as an added layer of validation, the authors had four experts evaluate a fraction of the dataset. In total, 2,800 samples were evaluated by an expert. Table 6.1 defines the recorded features for each cough sample as illustrated in (Orlandic et al., 2021).

*Table 6.1: The recorded features of cough samples in coughvid. Those features become the columns of the metadata table.*

| Name | Mandatory | Range of possible values | Description |
|------|-----------|--------------------------|-------------|
| uuid | Yes | String value | A unique id for each coughs ample |
| datetime | Yes | UTC date and time in ISO 8601 format | Timestamp of the received recording. |
| cough_detected | Yes | Floating point in the interval [0, 1] | Probability that the recording contains cough sounds, according to the automatic detection algorithm described in the Methods section. |
| SNR | Yes | Floating point in the interval $[0, \infty)$ | An estimation of the signal-to-noise ratio of the cough recording. |
| latitude | No | Floating point value | Self-reported latitude geolocation coordinate with reduced precision. |
| longitude | No | Floating point value | Self-reported longitude geolocation coordinate with reduced precision. |
| age | No | Integer value | Self-reported age value. |
| gender | No | {female, male, other} | Self-reported gender. |
| respiratory_condition | No | {True, False} | The patient has other respiratory conditions (self-reported). |
| fever_muscle_pain | No | {True, False} | The patient has fever or muscle pain (self-reported). |
| status | No | {COVID, symptomatic, healthy} | The patient self-reports that has been diagnosed with COVID-19 (COVID), that has symptoms but no diagnosis (symptomatic), or that is healthy (healthy). |
| expert_labels_{1,2,3,4} | No | JSON dictionary with the manual labels from experts 1, 2, 3 or 4 | The expert annotation variables are described in Table 4. |

## 7 Exploratory Data Analysis

This section analyzes the metadata and structure of the audio samples to validate the distribution and explore patterns. Ultimately, the goal of the project is to aid and improve the covid-19 detection through coughs. Thus, the goal of exploratory data analysis is to understand the trends in the dataset and identify potential biases. The metadata is loaded and the first 3 rows of the data are displayed, as shown in figure 7.1, for inspection of the metadata as shown in table 7.1.

*Figure 7.1: Loading the metadata of the coughvid dataset*

```python
# Load the metadata info
data_raw = pd.read_csv(os.path.join(ROOT, "metadata_compiled.csv"))
data_raw.head(3)
```

*Table 7.1: the first 3 rows of the coughvid metadata*

| uuid | datetime | cough_detected | SNR | latitude | longitude | age | gender | respiratory_condition | fever_muscle_pain | ... | quality_4 |
|------|----------|----------------|-----|----------|-----------|-----|--------|----------------------|-------------------|-----|-----------|
| 00039425-7f3a-42aa-ac13-834aaa2b6b92 | 2020-04-13T21:30:59.801831+00:00 | 0.9609 | 16.151433 | 31.3 | 34.8 | 15.0 | male | False | False | ... | NaN |
| 0007c6f1-5441-40e6-9aaf-a761d8f2da3b | 2020-10-18T15:38:38.205870+00:00 | 0.1643 | 16.217201 | NaN | NaN | 46.0 | female | False | False | ... | NaN |
| 0009eb28-d8be-4dc1-92bb-907e53bc5c7a | 2020-04-12T04:02:18.159383+00:00 | 0.9301 | 20.146058 | 40.0 | -75.1 | 34.0 | male | True | False | ... | NaN |

ws × 51 columns

The metadata contains 51 columns which were difficult to include in the table. Figure 6.2 shows all of the columns of the metadata.

*Figure 7.2: Coughvid metadata columns*

```
Index(['uuid', 'datetime', 'cough_detected', 'SNR', 'latitude', 'longitude',
'age', 'gender', 'respiratory_condition', 'fever_muscle_pain', 'status',
'quality_1', 'cough_type_1', 'dyspnea_1', 'wheezing_1', 'stridor_1', 'choking_1',
'congestion_1', 'nothing_1', 'diagnosis_1', 'severity_1', 'quality_2',
'cough_type_2', 'dyspnea_2', 'wheezing_2', 'stridor_2', 'choking_2',
'congestion_2', 'nothing_2', 'diagnosis_2', 'severity_2', 'quality_3',
'cough_type_3', 'dyspnea_3', 'wheezing_3', 'stridor_3', 'choking_3',
'congestion_3', 'nothing_3', 'diagnosis_3', 'severity_3', 'quality_4',
'cough_type_4', 'dyspnea_4', 'wheezing_4', 'stridor_4', 'choking_4',
'congestion_4', 'nothing_4', 'diagnosis_4', 'severity_4'], dtype='object')
```

Inspecting the columns shows a distinction between columns where the first couple are metadata associated with the audio such as its id, datetime, condition, etc. Those columns and their significance are summarized in table 5.1. Then there is a set of duplicated columns where each column is followed by a number between 1 and 4 such as diagnosis which is present four times in the columns `diagnosis_1`, `diagnosis_2`, `diagnosis_3`, and `diagnosis_4`. Those columns represent one of four experts' feedback on the feature as recorded. For example, `diagnosis_1` and `'severity_1'` represent the diagnosis and severity measured by expert 1 respectively. A row with null values for those columns means that the respective expert was not assigned the instance. Table 7.2 summarizes and describes each feature recorded by the experts. According to the dataset documentation, each of the four experts was assigned a unique set of audio samples to inspect; however, the four experts did not cover the entire dataset. Thus before modifying the structure of the metadata to reflect a better representation of the experts' feedback, exploratory data analysis is applied to the non-expert related features or columns.

*Table 7.2: Expert labeling description and range of possibilities. As stated each of the following features is present in the metadata table four times each time followed by the index of the expert responsible for the evaluation.*

| Name | Range of possible values | Description |
|---|---|---|
| **Quality** | {good, ok, poor, no_cough} | Quality of the recorded cough sound. |
| **cough_type** | {wet, dry, unknown} | Type of cough. |
| **dyspnea** | {True, False} | Audible dyspnea. |
| **wheezing** | {True, False} | Audible wheezing. |
| **Stridor** | {True, False} | Audible stridor. |
| **choking** | {True, False} | Audible choking. |
| **congestion** | {True, False} | Audible nasal congestion. |
| **nothing** | {True, False} | Nothing specific is audible. |
| **diagnosis** | {upper_infection, lower_infection, obstructive_disease, COVID-19, healthy_cough} | Impression of the expert about the condition of the patient. It can be an upper or lower respiratory tract infection, an obstructive disease (Asthma, COPD, etc), COVID-19, or a healthy cough. |
| **severity** | {pseudocough, mild, severe, unknown} | Impression of the expert about the severity of the cough. It can be a pseudocough from a healthy patient, a mild or severe cough from a sick patient, or unknown if the expert can't tell. |

## 7.1   Initial data cleaning

The first step in exploratory data analysis is exporting the percentage of null values per column. This allows for informed processing of data and a set of expectations. Figure 7.3 is used to generate table 7.3.

*Figure 7.3: Evaluating the null percentage of the columns in the metadata table*

```
# Check the column level Null %
# Check data Type of each of the column
temp_df = pd.DataFrame((data_raw.isnull().sum() / data_raw.shape[0]) * 100,
columns=["Null value%"])
temp_df["data_type"] = data_raw.dtypes
temp_df
```

*Table 7.3: Null value percentage for the metadata columns*

|  | Null value% | data_type |
|---|---|---|
| **uuid** | 0.000000 | object |
| **datetime** | 0.000000 | object |
| **cough_detected** | 0.000000 | float64 |
| **SNR** | 0.000000 | float64 |
| **latitude** | 41.620385 | float64 |
| **longitude** | 41.620385 | float64 |
| **age** | 44.760245 | float64 |
| **gender** | 41.108570 | object |
| **respiratory_condition** | 41.108570 | object |
| **fever_muscle_pain** | 41.108570 | object |
| **status** | 41.108570 | object |
| **quality_1** | 97.023485 | object |
| **cough_type_1** | 97.088824 | object |
| **dyspnea_1** | 97.023485 | object |
| **wheezing_1** | 97.023485 | object |
| **stridor_1** | 97.023485 | object |
| **choking_1** | 97.023485 | object |
| **congestion_1** | 97.023485 | object |
| **nothing_1** | 97.023485 | object |
| **diagnosis_1** | 97.088824 | object |
| **severity_1** | 97.088824 | object |
| **quality_2** | 97.023485 | object |
| **cough_type_2** | 97.092453 | object |
| **dyspnea_2** | 97.023485 | object |
| **wheezing_2** | 97.023485 | object |
| **stridor_2** | 97.023485 | object |
| **choking_2** | 97.023485 | object |
| **congestion_2** | 97.023485 | object |
| **nothing_2** | 97.023485 | object |
| **diagnosis_2** | 97.088824 | object |
| **severity_2** | 97.088824 | object |
| **quality_3** | 97.023485 | object |
| **cough_type_3** | 97.114233 | object |
| **dyspnea_3** | 97.023485 | object |
| **wheezing_3** | 97.023485 | object |
| **stridor_3** | 97.023485 | object |
| **choking_3** | 97.023485 | object |
| **congestion_3** | 97.023485 | object |
| **nothing_3** | 97.023485 | object |
| **diagnosis_3** | 97.125123 | object |
| **severity_3** | 97.114233 | object |
| **quality_4** | 97.023485 | object |
| **cough_type_4** | 97.092453 | object |
| **dyspnea_4** | 97.023485 | object |
| **wheezing_4** | 97.023485 | object |
| **stridor_4** | 97.023485 | object |
| **choking_4** | 97.023485 | object |
| **congestion_4** | 97.023485 | object |
| **nothing_4** | 97.023485 | object |
| **diagnosis_4** | 97.136012 | object |
| **severity_4** | 97.099713 | object |

Inspecting table 7.3 it is noted that only uuid, datetime, cough_detected, and SNR have no null values. Those columns are reliable in analysis. Furthermore, the remaining non-expert features - latitude, longitude, age, gender, respiratory_condition, fever_muscle_pain, and status – have around 41% null values which might distort the evaluation and findings; however, since the goal of the exploratory data analysis is to understand the data and detect potential biases this percentage of loss is acceptable. The null values will be replaced with "unknown" to reflect on the plots to be generated using the snippet shown in figure 7.4. Finally, the expert features had a 97% null percentage which is to be expected given that each expert evaluated a fraction of 2,800 samples out of the whole 25,000 samples. In total, experts have evaluated 12% of the dataset which means that their evaluation cannot be generalized or used without significantly reducing the dataset.

*Figure 7.4: Filling missing values in metadata with "unknown"*

```
# Handle the missing values
data_full = data_raw.fillna("unknown")
```

## 7.2   Data analysis

After the data is cleaned, the next step is to explore the data distribution and identify potential biases in the coughs. The findings of the exploratory data analysis will guide the selection and preprocessing stage of the pipeline.

### 7.2.1   Data Distribution Biases

Figure 7.5 plots the data distributions with respect to the status of the patient, the presence of respiratory conditions, gender, and the presence of fever or muscle pain.

*Figure 7.5: Plotting the data distribution of the coughvid*

```
f, axs = plt.subplots(2, 2, figsize=(12, 4), gridspec_kw={"width_ratios": [4, 4]})
sns.countplot(data=data_full, x="gender", ax=axs[0, 0])
sns.countplot(data=data_full, x="status", ax=axs[0, 1])
sns.countplot(data=data_full, x="respiratory_condition", ax=axs[1, 0])
sns.countplot(data=data_full, x="fever_muscle_pain", ax=axs[1, 1])
f.tight_layout()
```

*Figure 7.6: Data distributions of the status of the patient, the presence of respiratory conditions, the gender, and the presence of fever muscle pain within coughvid*

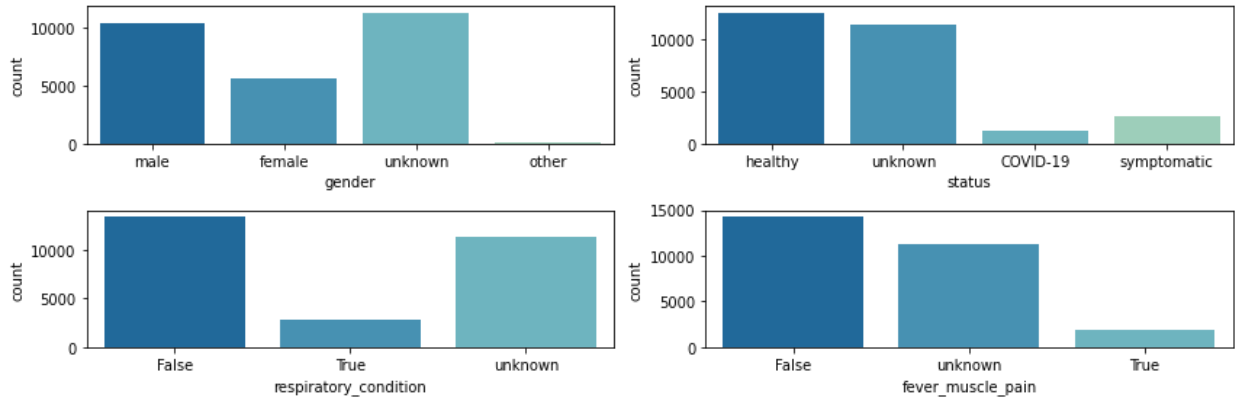

The data distribution shows for the most part that many participants chosen to not fill out many of the optional fields which is a consequence of the open voluenteering participation. Examning the first plot a clear bias towards males is present as the number of male particpants is twice the female. The second plot shows the status distribution of the patient which highlights the scarcity of covid samples as only around 1000 samples are present compared to other status with around tens of thousand samples. Respiratory condition shows that most participants have no respiratory conditions. Likewise, the majority of participants had no fever or muscle pain.

### 7.2.2   Quality examination

Two columns are examined to assess the quality of individual audio samples: cough_detected, and SNR. The former represents the probability that the audio sample contains a cough and examining the distribution of the feature shows a high likely hood of non-cough audio to be present as shown in figure 7.7.

*Figure 7.7: Cough detected random variable distribution*

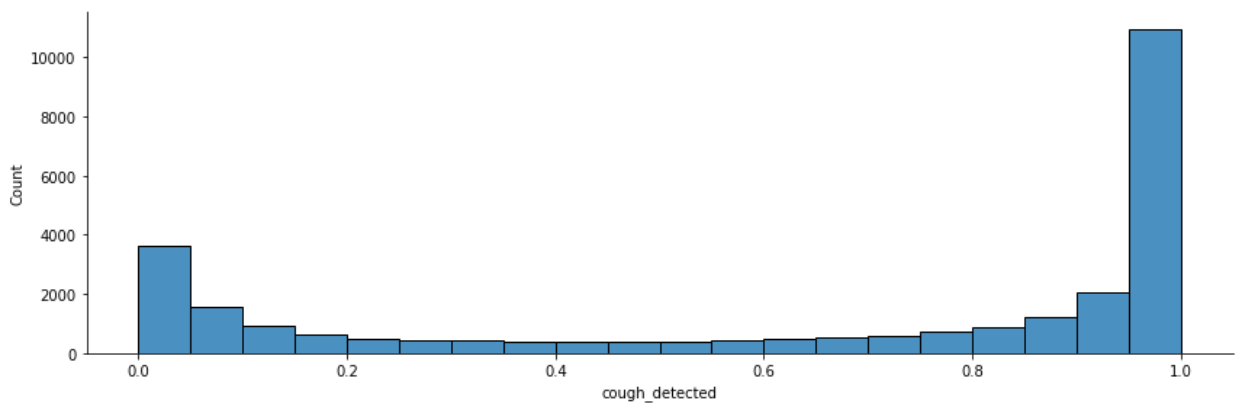

Thus, as shown in figure 7.9 only audio samples with a cough detected value of 0.5 or higher will be selected for the task to mitigate the presence of none cough audio. SNR or Signal to Noise Ratio is an estimation of the signal-to-noise ratio of the cough recording. Figure 7.8 shows the value distribution of the SNR feature.

*Figure 7.8: the value distribution of the SNR feature*

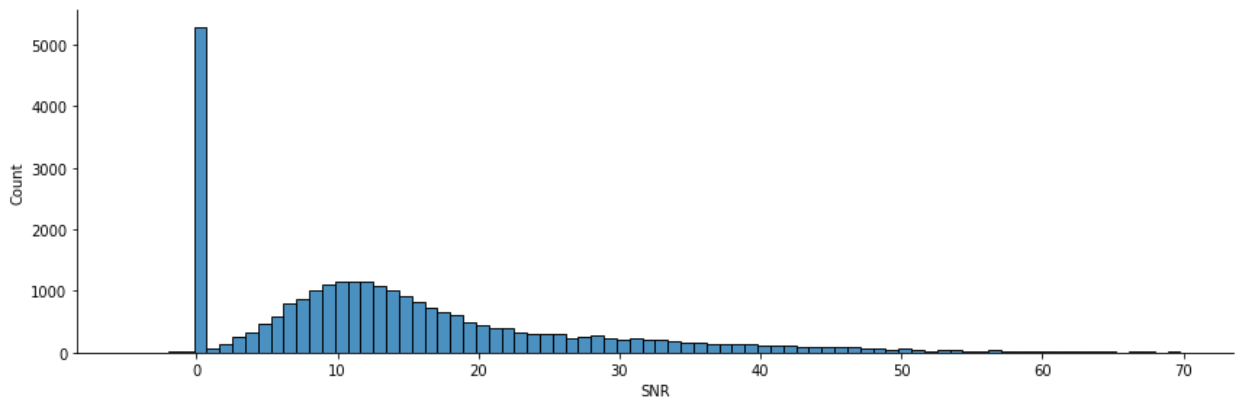

Figure 7.8 shows a huge spike of values with a very low SNR followed by a reasonable distribution of SNR values. To ensure that noise does not sabotage the results of training only samples with an SNR value greater than 5 are selected as shown in figure 7.9.

*Figure 7.9: Selecting audio samples with detected cough probability of 0.5 or higher and SNR value of 5 or greater*

```
filtered_data = data_full[(data_full["cough_detected"] > 0.5) & (data_full["SNR"]
> 5)]
```

The resulting distribusions after filtering the data are shown in figure 7.10.

*Figure 7.10: Data distribusion after filteration*

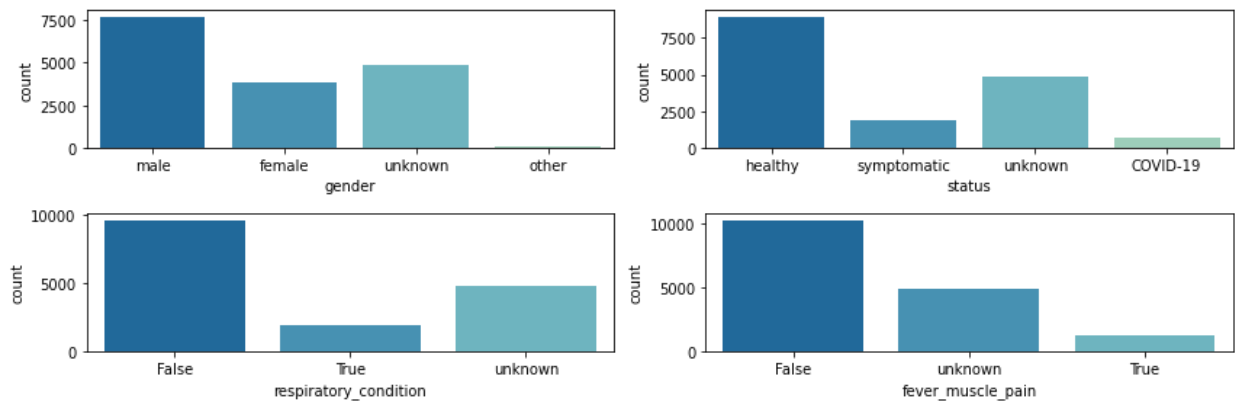

## 7.3 Expert annotation

While expert annotation was provided for only 2,800 samples, it is still worth examining the findings to understand the samples. The function defined in figure 7.11 is designed to take in a data frame, remove the 40 columns where each 10 represents the findings of an expert shown in table 7.2, and compile the 40 columns into 10 columns adding a physician column to represent the expert responsible for the findings. The function is then used in figure 7.12 to process the filtered data.

*Figure 7.11: Split by physicians, a function to remove the 40 columns of expert annotation and transform them into 11 where 10 represents the findings of the physician and one represents the physicians responsible for the findings.*

```python
def split_by_physicians(df):
    column_names = [
        "uuid",
        "datetime",
        "cough_detected",
        "SNR",
        "latitude",
        "longitude",
        "age",
        "gender",
        "respiratory_condition",
        "fever_muscle_pain",
        "status",
        "quality",
        "cough_type",
        "dyspnea",
        "wheezing",
        "stridor",
        "choking",
        "congestion",
        "nothing",
        "diagnosis",
        "severity",
    ]
    physician_01 = df.iloc[:, 0:21]
    physician_01 =
physician_01[physician_01.quality_1.notna()].reset_index(drop=True)
    physician_01.columns = column_names
    physician_01["physician"] = "P01"

    physician_02 = pd.concat([df.iloc[:, 0:11], df.iloc[:, 21:31]], axis=1)
    physician_02 =
physician_02[physician_02.quality_2.notna()].reset_index(drop=True)
    physician_02.columns = column_names
    physician_02["physician"] = "P02"

    physician_03 = pd.concat([df.iloc[:, 0:11], df.iloc[:, 31:41]], axis=1)
    physician_03 =
physician_03[physician_03.quality_3.notna()].reset_index(drop=True)
    physician_03.columns = column_names
```

```
    physician_03["physician"] = "P03"

    physician_04 = pd.concat([df.iloc[:, 0:11], df.iloc[:, 41:51]], axis=1)
    physician_04 =
physician_04[physician_04.quality_4.notna()].reset_index(drop=True)
    physician_04.columns = column_names
    physician_04["physician"] = "P04"
    return physician_01, physician_02, physician_03, physician_04
```

*Figure 7.12: Using the split by physician function to process the filtered dataset*

```
physician_01, physician_02, physician_03, physician_04 = split_by_physicians(filtered_data)
annotated_df = pd.concat(
    [physician_01, physician_02, physician_03, physician_04]
).reset_index(drop=True)
annotated_df = annotated_df.fillna("unknown")
annotated_df.head(3)
```

### 7.3.1 Physician distribution

The following figure shows the physician distribution by status, where overall as the four physicians were assigned the same number of samples with diverse statuses. The number of unknown statuses handled by physicians was minimal with physician four having noticeably a lower count.

*Figure 7.13: The right side shows the sample count by the physician and the left is for the samples handled by physicians based on the status of the patient.*

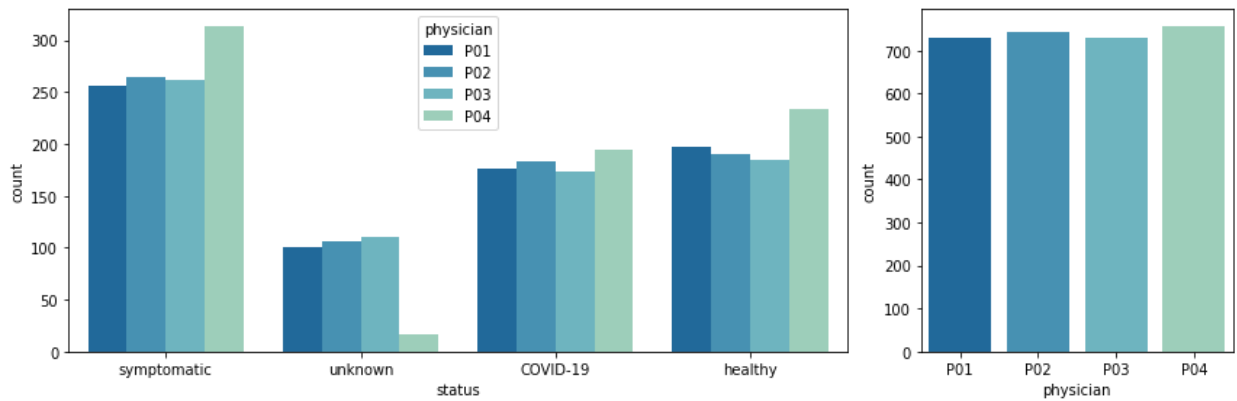

### 7.3.2 Cough quality

Physicians assessed the cough quality of their assigned samples. The results are shown in figure 7.14. Overall the majority of samples had a good quality and inspecting the cough quality relative to the status of the patient shows no biases of cough quality against a specific status as shown in figure 7.14.

*Figure 7.14: The left plot shows the overall cough quality distribution. The right plot shows the cough quality relative to the status of the patient*

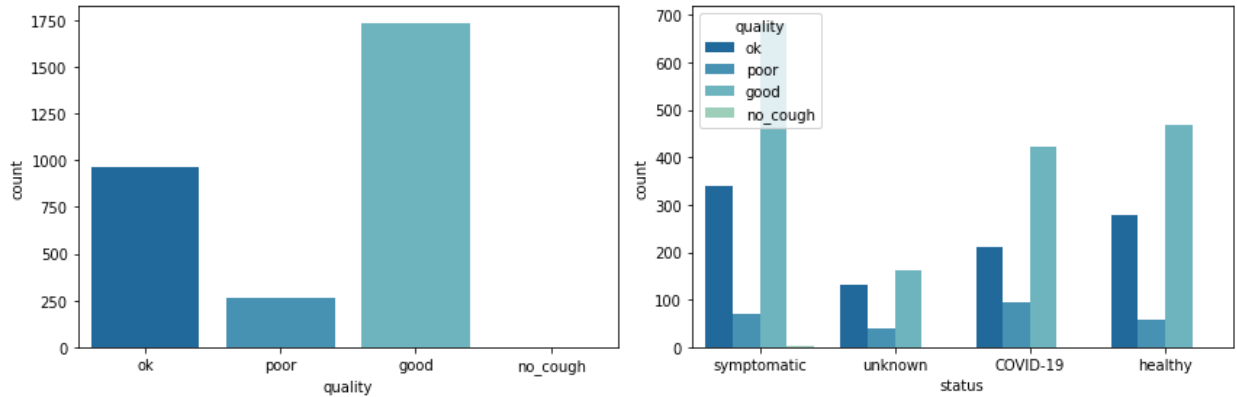

### 7.3.3 Cough type

The physicians also assess the cough type of each sample, labeling the cough as dry, wet, or unknown. Figure 7.15 shows the overall cough type distribution and the distribution relative to the patient's status. The most common cough type is the dry cough which is the common cough type present in covid-19 infected patients (Diwan, 2020).

*Figure 7.15: Left. Overall cough type distribution. Right. The cough type distribution relative to the patient's status*

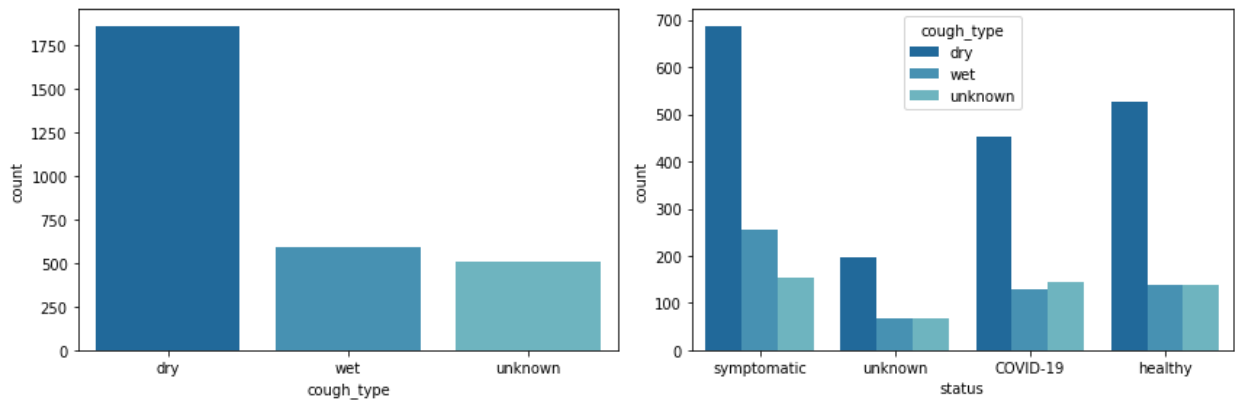

### 7.3.4 Cough diagnosis and severity

The physicians also analyzed the cough samples and provided a diagnosis of each along with the severity of the cough as shown in figure 7.16. While all data samples were labeled as healthy, symptomatic, covid-19 infected, or unknown - the physicians labeled the coughs as healthy, lower infection, covid-19 infected obstructive disease, upper infection, or unknown. The overall severity of the patient's cases was mild and as shown in figure 7.17 the severity with respect to the diagnosis shows no anomalies.

*Figure 7.16: Left. Cough diagnosis. Right. Cough severity*

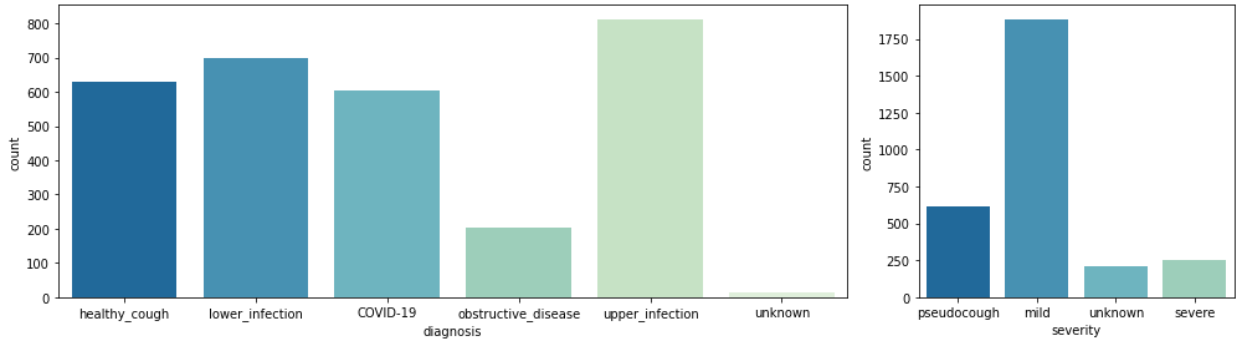

*Figure 7.17: Physicians' diagnosis with respect to the severity of the patient's case.*

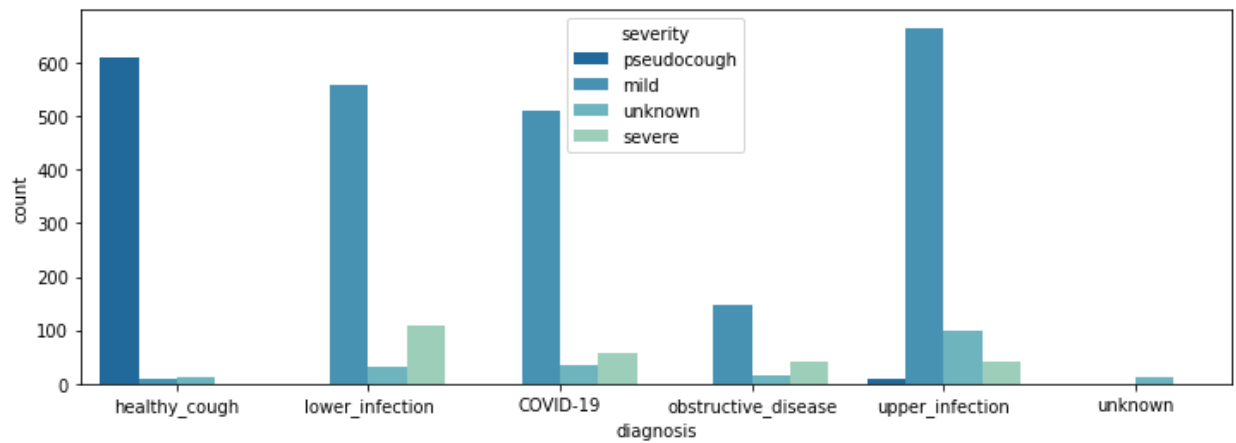

### 7.3.5   Expert status annotation vs patient diagnosis

Figure 7.18 shows the diagnosis and severity of the cough with respect to the physicians, where it can be noted that there is a difference between the given status to the physicians, shown in figure 7.13, and the physicians' diagnosis.

*Figure 7.18: Left. Cough diagnosis with respect to the physicians. Right. Cough severity with respect to the physicians.*

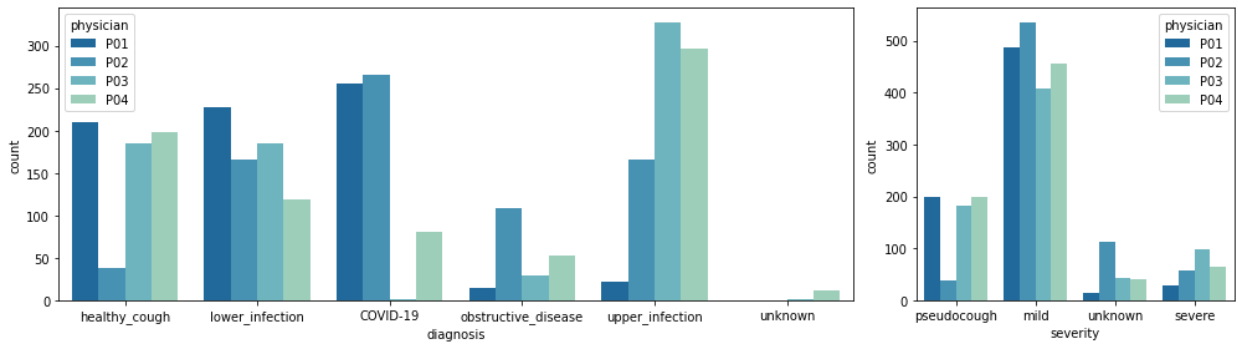

Examine the diagnosis match using the snippet shown in figure 7.19, the diagnosis to status match is presented in figure 7.20.

*Figure 7.19: Code snippet for representing the diagnosis to status match*

```python
def get_diagnosis_match(uuid):
    row = annotated_df[annotated_df["uuid"] == uuid].iloc[0]

    if row.status == "healthy":
        return row.diagnosis == "healthy_cough"
    elif row.status == "symptomatic":
        return row.diagnosis in [
            "lower_infection","upper_infection","obstructive_disease"
        ]
    elif row.status == "COVID-19":
        return row.diagnosis == "COVID-19"
    elif row.status == "unknown":
        return row.diagnosis == "unknown"

annotated_df["diagnosis_match"] = annotated_df["uuid"].apply(get_diagnosis_match)

plt.figure(figsize=(12, 4))
sns.countplot(data=annotated_df, x="status", hue="diagnosis_match")
plt.show()
```

*Figure 7.20: The diagnosis to status match*

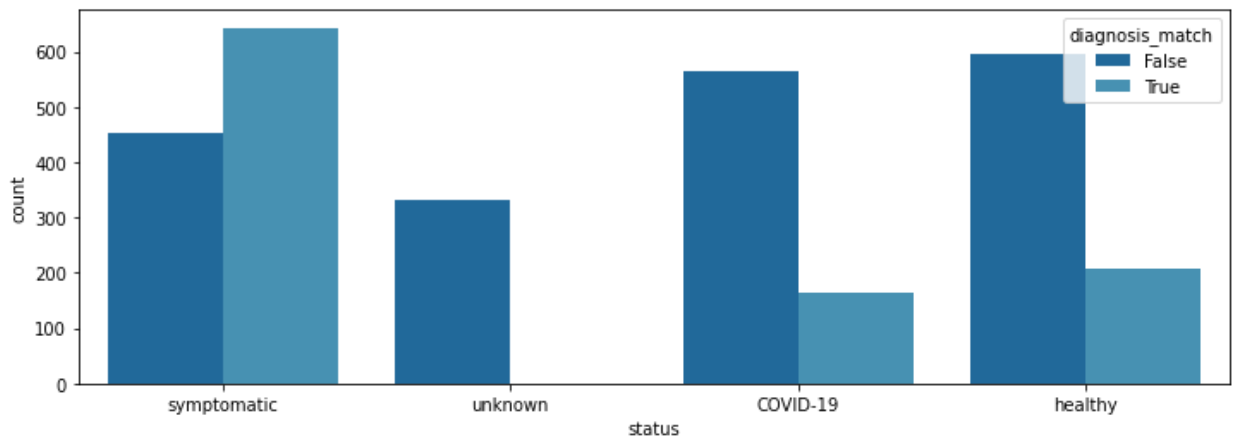

The findings show that the physicians' diagnosis is vastly different from the patient's reported status. Those findings are not disqualifying but certainly, raise doubt on the validity of the reported status by the patients.

### 7.3.6   Audible annotations

Figure 7.21 shows the presence of audible dyspnea, wheezing, choking, or congestion. Overall, very few samples contain any of that audible condition and they will be dropped so as not to affect the training process of the model.

*Figure 7.21: the presence of audible dyspnea, wheezing, choking, or congestion in the dataset*

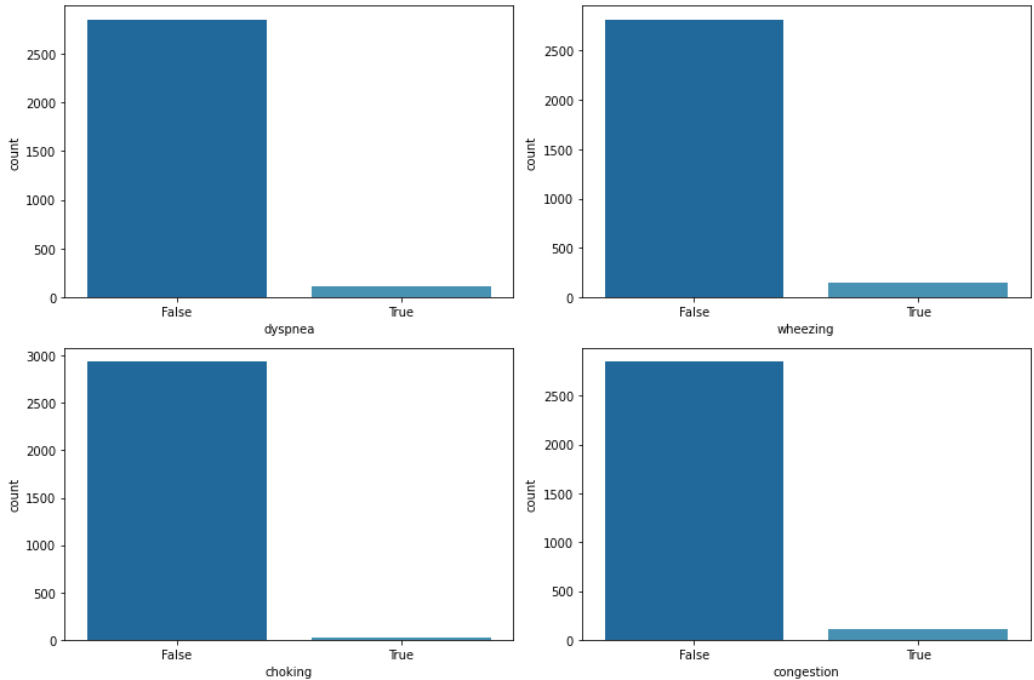

*Figure 7.22: the presence of audible dyspnea, wheezing, choking, or congestion in the dataset with respect to the cough status*

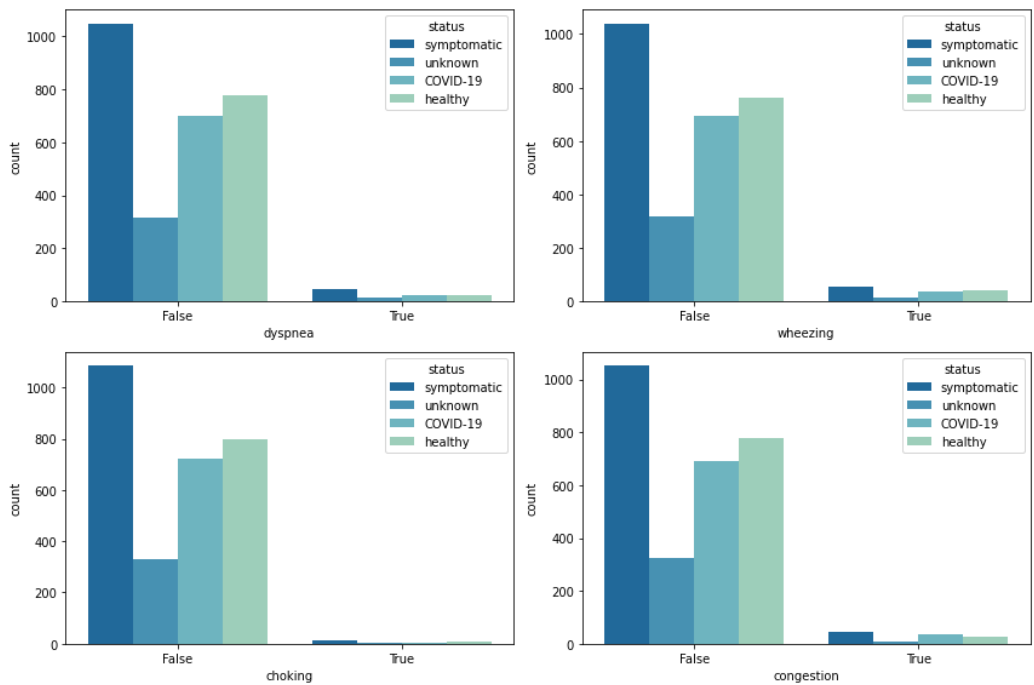

# 8    Audio pre-processing

After exploring the data and cleaning it appropriately the next step is to preprocess the audio data. Figure 8.1 shows a randomly selected sample from each of the three classes of audio: healthy, symptomatic, and covid-19.

*Figure 8.1: Randomly selected sample from each of the three classes of audio: healthy, symptomatic, and covid-19.*

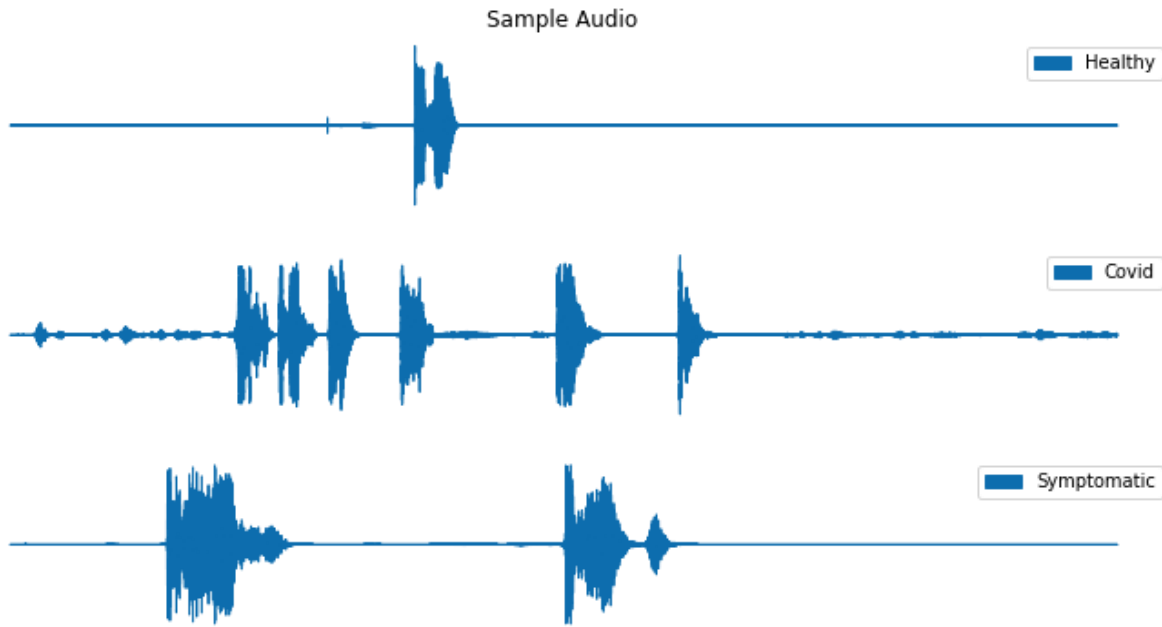

Examining the samples, it can be noted that audio samples vary in length and have a lot of silence or noise. Thus, the preprocessing step must not only normalize and prepare the audio for the model but also uniform the audio duration and handle noise and silences.

## *8.1    Preprocessing audio data*

The function shown in figure 8.2 loads an audio sample and normalizes and downsamples the audio for better representation. The function normalizes the audio segment by converting the audio to mono, or a single audio channel. Then the audio values are normalized into a range of -1 to 1. Lastly, the audio is downsampled for antialiasing.

*Figure 8.2: A function to load and preprocess the cough audio samples*

```
from scipy import signal
from scipy.signal import butter, filtfilt

def preprocess_cough(x, fs, cutoff=6000, normalize=True, filter_=True,
downsample=True):
```

```python
    # Normalize, lowpass filter, and downsample cough samples in a given data
folder
    fs_downsample = cutoff * 2

    # Preprocess Data
    if len(x.shape) > 1:
        x = np.mean(x, axis=1)  # Convert to mono
    if normalize:
        x = x / (np.max(np.abs(x)) + 1e-17)  # Norm to range between -1 to 1
    if filter_:
        b, a = butter(
            4, fs_downsample / fs, btype="lowpass"
        )  # 4th order butter lowpass filter
        x = filtfilt(b, a, x)
    if downsample:
        x = signal.decimate(x, int(fs / fs_downsample))  # Downsample for anti-
aliasing

    fs_new = fs_downsample

    return np.float32(x), fs_new
```

## 8.2   Cough segmentation for data augmentation

Reexamining the samples shown in figure 8.1 it can be noted that a single audio sample contains multiple cough instances; therefore, a single audio sample can be segmented into multiple samples and normalized to have a uniform length. The function shown in figure 8.3 segments a given processed cough into sample segments each containing a single cough. This is done by setting a threshold for the single length where any single length below the threshold is audio, then the consecutive signals above the threshold are concatenated into a single audio segment, then the process is repeated.

*Figure 8.3: A function for generating segmentation masks for coughs as well as the segmented cough segments*

```python
def segment_cough(
    x, fs, cough_padding=0.2, min_cough_len=0.2, th_l_multiplier=0.1,
th_h_multiplier=2
):
    # Preprocess the data by segmenting each file into individual coughs using a
hysteresis comparator on the signal power

    cough_mask = np.array([False] * len(x))

    # Define hysteresis thresholds
    rms = np.sqrt(np.mean(np.square(x)))
```

```python
        seg_th_l = th_l_multiplier * rms
        seg_th_h = th_h_multiplier * rms

        # Segment coughs
        cough_segments = []
        padding = round(fs * cough_padding)
        min_cough_samples = round(fs * min_cough_len)
        cough_start = 0
        cough_end = 0
        cough_in_progress = False
        tolerance = round(0.01 * fs)
        below_th_counter = 0

        for i, sample in enumerate(x ** 2):
            if cough_in_progress:
                if sample < seg_th_l:
                    below_th_counter += 1
                    if below_th_counter > tolerance:
                        cough_end = i + padding if (i + padding < len(x)) else len(x)
- 1

                        cough_in_progress = False
                        if cough_end + 1 - cough_start - 2 * padding >
min_cough_samples:
                            cough_segments.append(x[cough_start : cough_end + 1])
                            cough_mask[cough_start : cough_end + 1] = True
                elif i == (len(x) - 1):
                    cough_end = i
                    cough_in_progress = False
                    if cough_end + 1 - cough_start - 2 * padding > min_cough_samples:
                        cough_segments.append(x[cough_start : cough_end + 1])
                else:
                    below_th_counter = 0
            else:
                if sample > seg_th_h:
                    cough_start = i - padding if (i - padding >= 0) else 0
                    cough_in_progress = True

        return cough_segments, cough_mask
```

Using the two functions in figures 8.3 and 8.2 cough segments and masks are identified as shown in figure 8.4 and the respective cough segments are generated using the masks. A sample of the resulting segments is shown in figure 8.5. The code snippet that applied the two functions is shown in figure 8.6.

*Figure 8.4: Cough masks in a sampled audio*

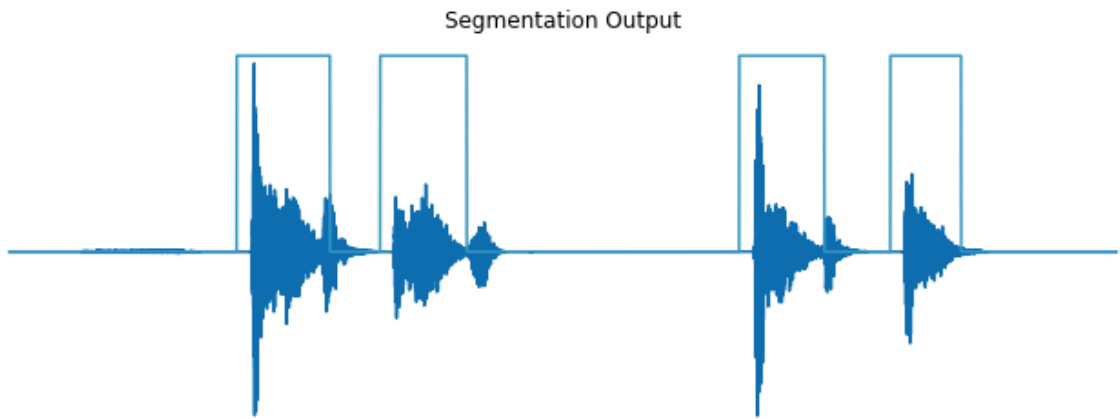

*Figure 8.5: generated cough segments of the audio sample shown in figure 8.4*

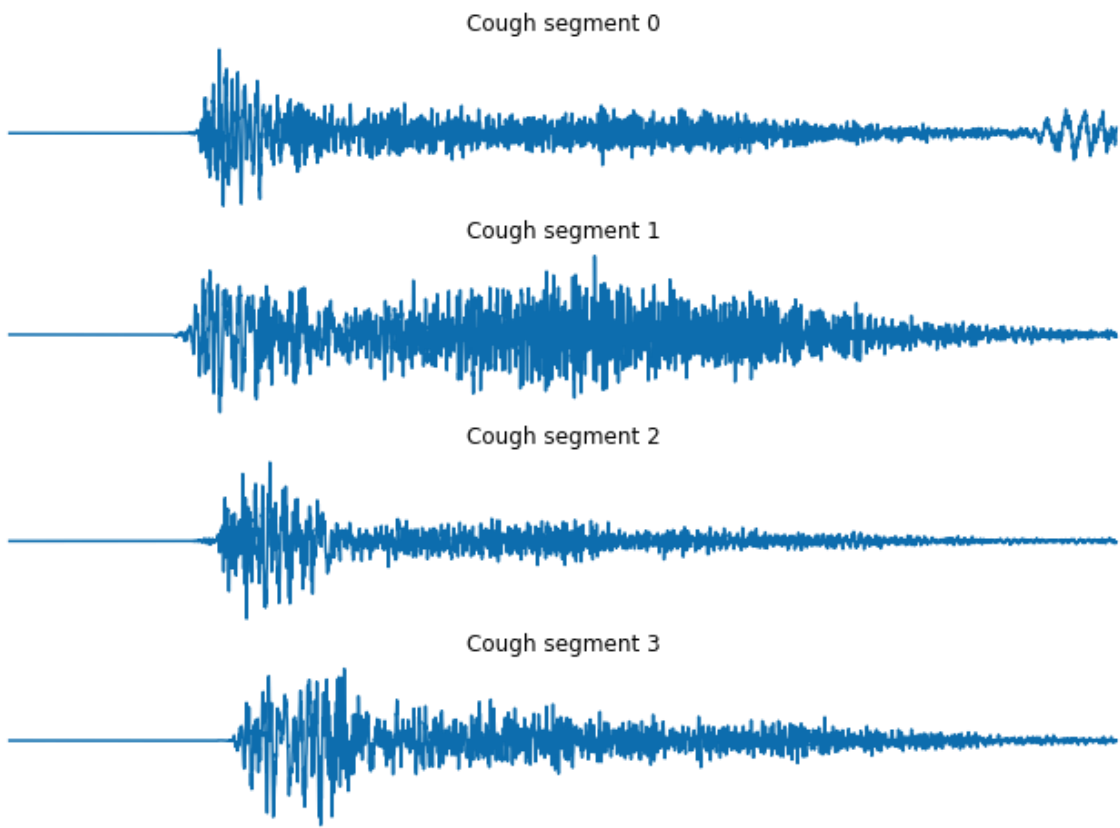

*Figure 8.6: Code snippet for processing an audio sample and generating the cough segments of the sample*

```
# Preprocess (Normalize, lowpass filter, and downsample cough samples)
processed_audio, sample_rate = preprocess_cough(healthy_audio, fs)
```

```python
# Segment each audio into individual coughs using a hysteresis comparator on the
signal power
cough_segments, cough_mask = segment_cough(
    processed_audio,
    sample_rate,
    min_cough_len=0.1,
    cough_padding=0.1,
    th_l_multiplier=0.1,
    th_h_multiplier=2,
)

fig = plt.figure(figsize=(12, 4))
plt.plot(processed_audio)
plt.plot(cough_mask)
plt.title("Segmentation Output")
plt.axis("off")
```

## 8.3   Storing the preprocess cough segments

The code snippet shown in figure 8.7 preprocesses a number of randomly selected samples from all of the classes. The selection is done such that all of the covid-19 audio samples are selected and an equal number of samples are randomly selected from each of the other classes. This selection approach balances out the data distribution by class. Then figure 8.8 shows a code snippet for storing the processed samples as NumPy files where a directory is created for each class. Saving the data as NumPy files makes the loading process significantly faster.

*Figure 8.7: Selecting an equal number of samples from each class where all of the covid samples are selected*

```python
class_size = len(data_full[data_full.status == "COVID-19"])
class2df = {}
for status in data_raw.status.dropna().unique():
    class2df[status] = data_full[data_full.status == status].sample(class_size)
selected_data = pd.concat(class2df.values())

audio_length_in_seconds = 1
for_saving = defaultdict(dict)
for i, row in tqdm(selected_data.iterrows()):
    audio_path = os.path.join(
        AUDIO_ROOT,
        row.uuid + ".wav",
    )
    audio, fs = librosa.load(audio_path, mono=True)

    # Preprocess (Normalize, lowpass filter, and downsample cough samples)
```

```python
    processed_audio, sample_rate = preprocess_cough(audio, fs)

    # Segment each audio into individual coughs using a hysteresis comparator on
the signal power
    cough_segments, cough_mask = segment_cough(
        processed_audio,
        sample_rate,
        min_cough_len=0.1,
        cough_padding=0.1,
        th_l_multiplier=0.1,
        th_h_multiplier=2,
    )

    cough_segments = [
        librosa.util.fix_length(
            cough_segment, int(sample_rate * audio_length_in_seconds)
        )
        for cough_segment in cough_segments
    ]

    coughs[row.status].extend(cough_segments)
    for_saving[row.status][row.uuid] = len(cough_segments)

coughs = {status: np.array(cough_segments) for status, cough_segments in
coughs.items()}
```

*Figure 8.8: Storing the preprocessed data into a NumPy file for fast processing and loading*

```python
if not os.path.exists(coughvid_cough_segments_path):
    os.makedirs(coughvid_cough_segments_path)

for status, uuid2cough_count in tqdm(for_saving.items()):
    status_path = os.path.join(coughvid_cough_segments_path, status.lower())
    if not os.path.exists(status_path):
        os.makedirs(status_path)

    saved_count = 0
    for uuid, cough_count in tqdm(uuid2cough_count.items()):
        cough_segments = coughs[status][saved_count : saved_count + cough_count]
        saved_count += cough_count

        for i, cough_segment in enumerate(cough_segments):
            np.save(
                os.path.join(status_path, f"{uuid}-{i}.npy"),
                cough_segment,
```

```
            allow_pickle=True,
)
```

## 9    Feature extraction

Following the preprocessing stage of the data comes feature extraction of the audio data. A common practice when developing a deep learning model for audio tasks is to represent the features of the audio visually and use the generated feature representation for the said task. It turns out that there are a couple of different algorithms for audio feature representation and selecting the optimal algorithm is key for the audio task. The following subsections discuss different feature extraction algorithms and will all be showing results performed on the randomly selected audio sample shown in figure 9.1.

*Figure 9.1: A randomly selected audio sample*

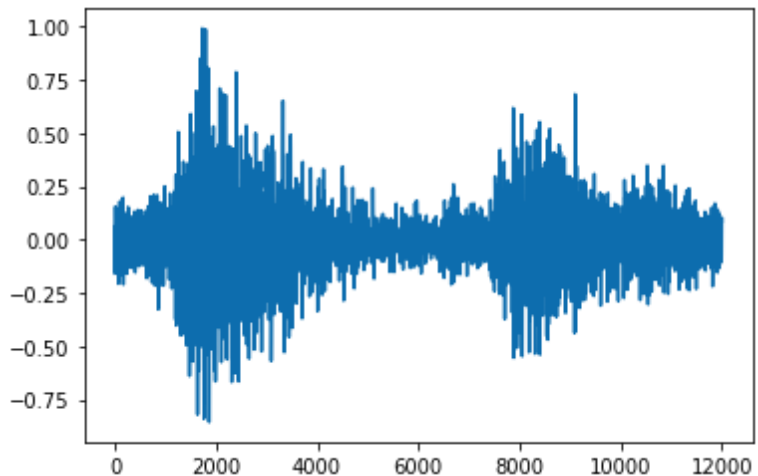

### 9.1    Spectrogram

A **spectrogram** is a graphical representation of the signal strength, or "loudness," of a signal over time at various frequencies present in a waveform. Not only can one see whether there is more or less energy at a particular frequency, such as 2 Hz vs 10 Hz, but also how the energy levels vary over time.

#### 9.1.1    Short-Time Fourier Transform

The Short-Time Fourier Transform (STFT) is an immensely useful tool for processing audio signals. It defines a particularly useful class of time-frequency distributions that specify the complex amplitude versus time and frequency characteristics of any signal. It is obtained by applying the Short-Time Fourier Transform (STFT) to the signal. In the simplest of terms, the STFT of a signal is calculated by applying the Fast Fourier Transform (FFT) locally on small-time segments of the signal. Typically, a spectrogram is depicted as a heat map, that is, an image with the intensity indicated by varying the color or brightness. Figure 9.2 shows the STFT of figure 9.1

*Figure 9.2: STFT of figure 9.1*

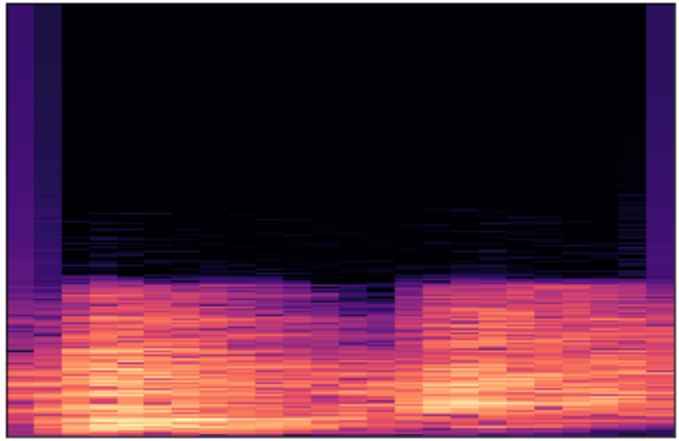

### 9.1.2   Mel-spectrogram

Humans perceive sound logarithmically. We are more sensitive to differences in lower frequencies than to differences in higher frequencies. For example, while we can easily distinguish between 500 and 1000 Hz, we will struggle to distinguish between 10,000 and 10,500 Hz, even if the distance between the two pairs is the same. As a result, the mel scale was created. It is a logarithmic scale based on the axiom that equal distances on the scale correspond to the same perceived distance. Figure 9.3 shows the mel-spectrogram of figure 9.1.

*Figure 9.3: the mel-spectrogram of figure 9.1*

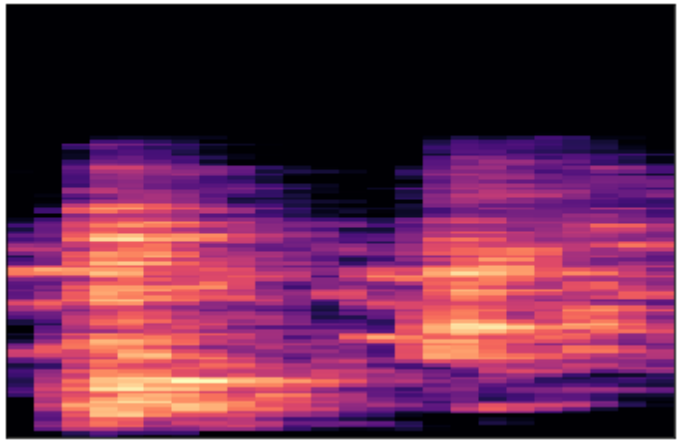

*Figure 9.4: the mel-spectrogram of figure 9.1 with axis and grey scaling*

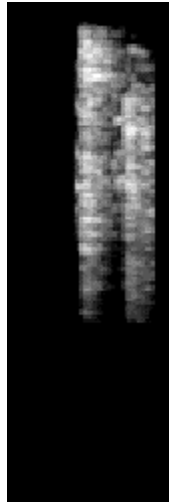

## 9.2   Mel-Frequency Cepstral Coefficients

The Mel-Frequency Cepstrum Coefficients (MFCCs) are the coefficients that comprise the mel-frequency cepstrum. The cepstrum of a signal contains information about the rate of change in its spectral bands.

A cepstrum is essentially a spectrum of the log of the time signal's spectrum. The resulting spectrum is neither in the frequency domain nor in the time domain and thus has been dubbed the quefrency domain. The cepstrum communicates the various values that contribute to the formation of a sound's formants (a characteristic component of the quality of a speech sound) and timbre. Figure 9.5 shows the MFCC of figure 9.1.

*Figure 9.5: MFCC of figure 9.1.*

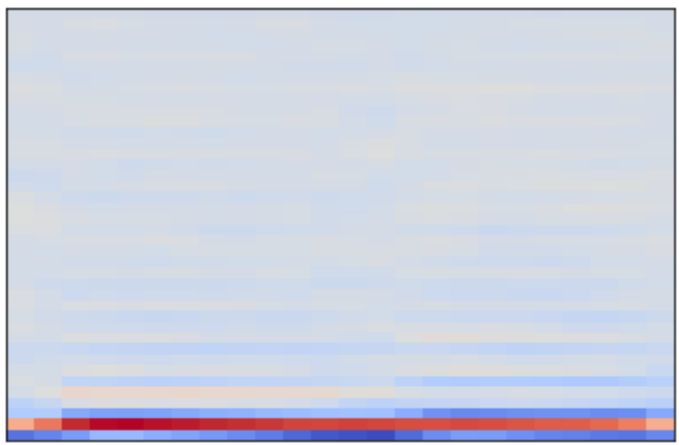

### 9.3   Chroma Features

Chroma features are a highly effective method of representing music audio in which we use a 12-element representation of spectral energy called a chroma vector in which each of the 12 bins represents one of the twelve equal-tempered pitch classes found in western-style music (semitone spacing). It can be calculated from the input sound signal's logarithmic short-time Fourier transform, also known as a chromatogram or pitch class profile. Figure 9.6 shows the chroma features of figure 9.1.

*Figure 9.6: chroma features of figure 9.1*

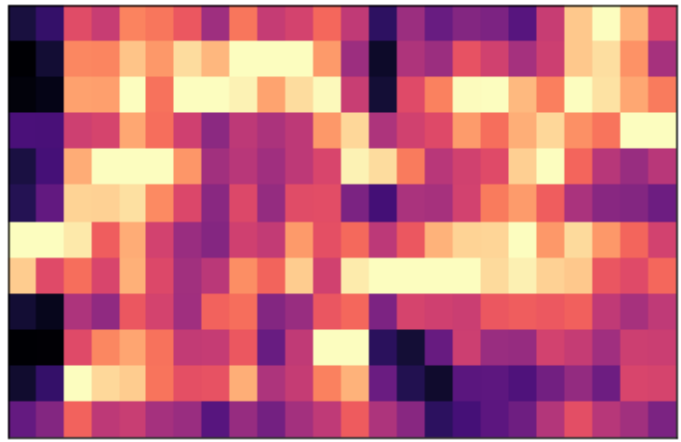

### 9.4   Selected feature representation algorithm

The selected feature representation algorithm is the mel-spectrogram as a representative algorithm for audio data. Mel-spectrogram offers a rich representation of the audio sample without being computationally expensive (Doshi, 2021).

## 10  Generative Adversarial Network Implementation

Generative adversarial networks have been discussed in detail in section 2.1 of this report. This section discusses the selected GAN pipeline and its implementation. Different GAN variations were tested for the task of cough synthesis including Deep Convolutional GAN (DCGAN), WaveGAN (Donahue et al., 2018), and Auxiliary Classifier GAN (ACGAN) (Odena et al., 2017). The ACGAN was selected as the best model for the task.

### 10.1  Auxiliary Classifier GAN (ACGAN)

Auxiliary Classifier GAN (ACGAN), shown in Figure 10.1, is a variation of cGAN proposed by (Odena et al., 2017) that alters the discriminator to predict the image class instead of receiving it as an input. The discriminator outputs two probability distributions, one representing whether the given image is real or generated and the other representing the class label of the given image. Likewise, the generator is given two inputs: the noise sampled from the latent space and the class label of the desired generated instance. This simple modification stabilizes the training.

*Figure 10.1: ACGAN architecture. c is the class label, z is the random noise, Q is the discriminator's probability distribution of the class labels, and y is the probability that the input to the discriminator is real*

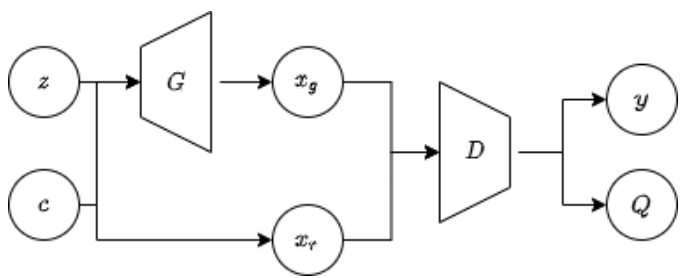

The selection of ACGAN for this task is optimal since it allows for the use of a diverse array of class labels for training the model. If a vanilla DCGAN or WaveGAN were used only the Covid-19 audio segment could be used for training the model; however, more class labels can be used to train the ACGAN since the generator is given the class label of the desired data.

### 10.2  Discriminator

The discriminator takes as input a spectrogram, which can be generated or real, and determines the class of the input as well as the probability of it being real. Figures 10.2 and 10.4 show the discriminator architecture for the ACGAN.

*Figure 10.2: discriminator architecture code*

```
def make_discriminator_model():
    model = models.Sequential()
    model.add(
```

```python
    layers.Conv2D(
        32,
        kernel_size=(3, 3),
        strides=(1, 1),
        input_shape=IMG_SHAPE,
        padding="same",
    )
)
model.add(layers.BatchNormalization(momentum=0))
model.add(layers.LeakyReLU(alpha=0.2))
model.add(layers.Dropout(0.5))

model.add(
    layers.Conv2D(
        64,
        kernel_size=(3, 3),
        strides=(2, 2),
        input_shape=IMG_SHAPE,
        padding="same",
    )
)
model.add(layers.BatchNormalization(momentum=0))
model.add(layers.LeakyReLU(alpha=0.2))
model.add(layers.Dropout(0.5))

model.add(layers.Conv2D(128, kernel_size=(3, 3), strides=(2, 2),
padding="same"))
model.add(layers.BatchNormalization(momentum=0))
model.add(layers.LeakyReLU(alpha=0.2))
model.add(layers.Dropout(0.5))

model.add(layers.Conv2D(256, kernel_size=(3, 3), strides=(2, 2),
padding="same"))
model.add(layers.BatchNormalization(momentum=0))
model.add(layers.LeakyReLU(alpha=0.2))
model.add(layers.Dropout(0.5))

model.add(layers.Conv2D(512, kernel_size=(3, 3), strides=(2, 2),
padding="same"))
model.add(layers.BatchNormalization(momentum=0))
model.add(layers.LeakyReLU(alpha=0.2))
model.add(layers.Dropout(0.5))

model.add(layers.Flatten())
```

```
img = layers.Input(shape=IMG_SHAPE)
features = model(img)

validity = layers.Dense(1, activation="sigmoid")(features)
label = layers.Dense(NUM_CLASSES, activation="softmax")(features)

tf.keras.utils.plot_model(model, "discriminator.png", show_shapes=True)
return models.Model(img, [validity, label])
```

### 10.2.1 Structure

The discriminator takes in a spectrogram of the uniformed shape and then processes it through a series of convolutional layers enhanced with batch normalization, leaky relu, and drop out.Then the output is flattened and passed to two different layers one for validation and one for labeling. The validity layer determines whether the given input is real or fake and uses a sigmoid activation function, while the label layer generated a probability distribution representing the likely hood of the given image belonging to a specific class and uses the softmax activation function.

### 10.2.2 Optimization

To train the model in a way that would optimize its ability to classify input and validate it two loss functions are used. First, the model is optimized using the Adam optimizer with a small learning rate which was proven to provide stable training. As for the loss functions both the ["binary_crossentropy", "sparse_categorical_crossentropy"] where selected. Binary cross-entropy optimizes the binary task of validating the realness of the input whereas sparse categorical cross entropy accurately measures the classification loss of the model.

*Figure 10.3: Optimizer and losses selection for the discriminator*

```
optimizer = tf.keras.optimizers.Adam(0.0002, 0.5)
losses = ["binary_crossentropy", "sparse_categorical_crossentropy"]

discriminator.compile(loss=losses, optimizer=optimizer, metrics=["accuracy"])
```

*Figure 10.4: discriminator architecture*

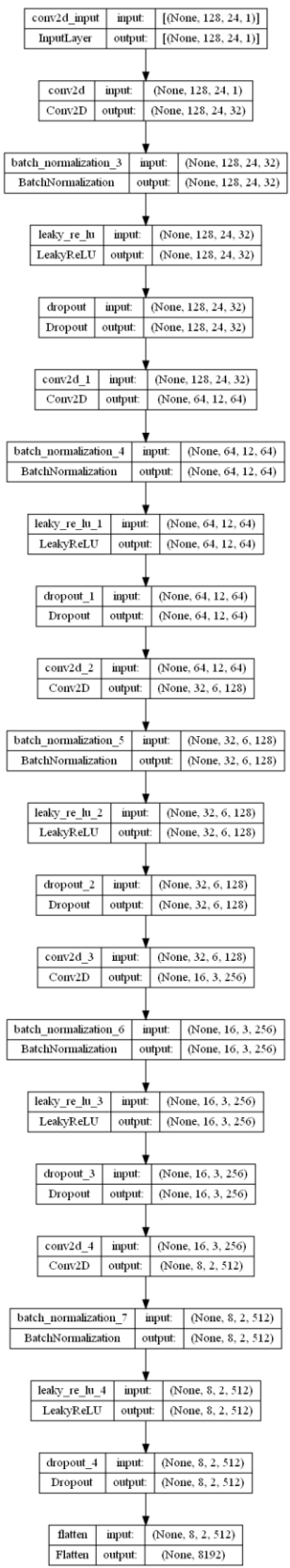

## 10.3  Generator

As stated the generator of the ACGAN network is modified to accept two inputs the noise and the class label of the data to be generated. Figure 10.5 and figure 10.9 show the architecture of the generator designed for generating cough audio segments.

*Figure 10.5: Generator architecture code*

```python
def make_generator_model():
    noise = layers.Input(shape=(LATENT_DIM,))
    label = layers.Input(shape=(1,), dtype="int32")

    noise_branch = layers.Dense(1024 * 16 * 3)(noise)
    noise_branch = activations.relu(noise_branch)
    noise_branch = layers.Reshape((16, 3, 1024))(noise_branch)
    noise_branch = models.Model(inputs=noise, outputs=noise_branch)

    label_branch = layers.Embedding(input_dim=50, output_dim=1)(label)
    label_branch = layers.Dense(7 * 7, input_shape=(7, 7))(label_branch)
    label_branch = activations.linear(label_branch)
    label_branch = layers.Reshape(
        (7, 7, 1),
    )(label_branch)
    label_branch = models.Model(inputs=label, outputs=label_branch)

    # weight initialization
    init = tf.keras.initializers.RandomNormal(stddev=0.02)

    combined = layers.concatenate([noise_branch.output, label_branch.output])

    combined = layers.Conv2DTranspose(
        512, (5, 5), strides=(2, 2), padding="same", kernel_initializer=init
    )(combined)
    combined = layers.BatchNormalization(momentum=0)(combined)
    combined = activations.relu(combined)

    combined = layers.Conv2DTranspose(
        256, (5, 5), strides=(2, 2), padding="same", kernel_initializer=init
    )(combined)
    combined = layers.BatchNormalization(momentum=0)(combined)
    combined = activations.relu(combined)

    combined = layers.Conv2DTranspose(
        128, (5, 5), strides=(1, 1), padding="same", kernel_initializer=init
    )(combined)
    combined = layers.BatchNormalization(momentum=0)(combined)
```

```
    combined = activations.relu(combined)

    combined = layers.Conv2DTranspose(
        1, (5, 5), strides=(2, 2), padding="same", kernel_initializer=init
    )(combined)
    combined = activations.tanh(combined)

    model = models.Model(
        inputs=[label_branch.input, noise_branch.input], outputs=combined
    )

    tf.keras.utils.plot_model(model, "generator.png", show_shapes=True)

    return model
```

### 10.3.1 Class label

While the ACGAN paper defined that the generator model should take a vector input that is a concatenation of the latent space and the class label, modern approaches separate the two into branches: the noise branch and the label branch. The label branch is perceived as an additional feature map which is achieved by using a learned embedding with an arbitrary number of dimensions, 50 in this case. The output of the embedding layer can then be passed to a fully connected layer with a linear activation function to generate a 7 by 7 feature map.

*Figure 10.6: The label branch*

```
    label_branch = layers.Embedding(input_dim=50, output_dim=1)(label)
    label_branch = layers.Dense(7 * 7, input_shape=(7, 7))(label_branch)
    label_branch = activations.linear(label_branch)
    label_branch = layers.Reshape(
        (7, 7, 1),
    )(label_branch)
    label_branch = models.Model(inputs=label, outputs=label_branch)
```

### 10.3.2 Noise

The noise from the defined latent space of 50,000 is interpreted by a fully connected layer with relu activation function to create 1024 16 by 3 feature maps, each feature map represents a low-resolution version of the expected output. The noise and label are then concatenated channel-wise before getting passed to the model.

*Figure 10.7: Noise branch and the concatenation of the noise and label branches*

```
    noise = layers.Input(shape=(LATENT_DIM,))
    label = layers.Input(shape=(1,), dtype="int32")
```

```python
    noise_branch = layers.Dense(1024 * 16 * 3)(noise)
noise_branch = activations.relu(noise_branch)
noise_branch = layers.Reshape((16, 3, 1024))(noise_branch)
noise_branch = models.Model(inputs=noise, outputs=noise_branch)

    …

    combined = layers.concatenate([noise_branch.output, label_branch.output])
```

### 10.3.3  2D transpose convolutional

Through a series of 2D transpose convolutional layers, the 16 by 3 feature map is upsampled by an iterative factor of two until the desired output shape of 128 by 24 is achieved. Relu activation functions are used in place of leakyRelu following the ACGAN paper. Finally using a tanh activation function the output of the generator is a single feature map with pixel values between [-1, 1].

*Figure 10.8: 2D convolutional transpose of the generator*

```python
# weight initialization
    init = tf.keras.initializers.RandomNormal(stddev=0.02)

    combined = layers.concatenate([noise_branch.output, label_branch.output])

    combined = layers.Conv2DTranspose(
        512, (5, 5), strides=(2, 2), padding="same", kernel_initializer=init
    )(combined)
    combined = layers.BatchNormalization(momentum=0)(combined)
    combined = activations.relu(combined)

    combined = layers.Conv2DTranspose(
        256, (5, 5), strides=(2, 2), padding="same", kernel_initializer=init
    )(combined)
    combined = layers.BatchNormalization(momentum=0)(combined)
    combined = activations.relu(combined)

    combined = layers.Conv2DTranspose(
        128, (5, 5), strides=(1, 1), padding="same", kernel_initializer=init
    )(combined)
    combined = layers.BatchNormalization(momentum=0)(combined)
    combined = activations.relu(combined)

    combined = layers.Conv2DTranspose(
        1, (5, 5), strides=(2, 2), padding="same", kernel_initializer=init
    )(combined)
```

```
    combined = activations.tanh(combined)

    model = models.Model(
        inputs=[label_branch.input, noise_branch.input], outputs=combined
    )

    tf.keras.utils.plot_model(model, "generator.png", show_shapes=True)

    return model
```

### 10.3.4  Optimization and loss

The generator network is optimized using the loss of the discriminator. This is achieved by creating a composite model comprising both the generator and discriminator such that the weights of the discriminator are not updated during the backpropagation of the composite model. What this achieves is an indirect way to optimize the generator. Since the composite model uses the discriminator loss it will be compiled in the same way as the discriminator.

*Figure 10.9: Constructing the composite model*

```
noise = layers.Input(shape=(LATENT_DIM,))
label = layers.Input(shape=(1,))
img = generator([label, noise])

discriminator.trainable = False
for layer in discriminator.layers:
    if not isinstance(layer, layers.BatchNormalization):
        layer.trainable = False

valid, target_label = discriminator(img)

combined = models.Model([label, noise], [valid, target_label])

combined.compile(loss=losses, optimizer=optimizer)
```

### 10.3.5  Generator architecture

*Figure 10.10: Generator architecture*

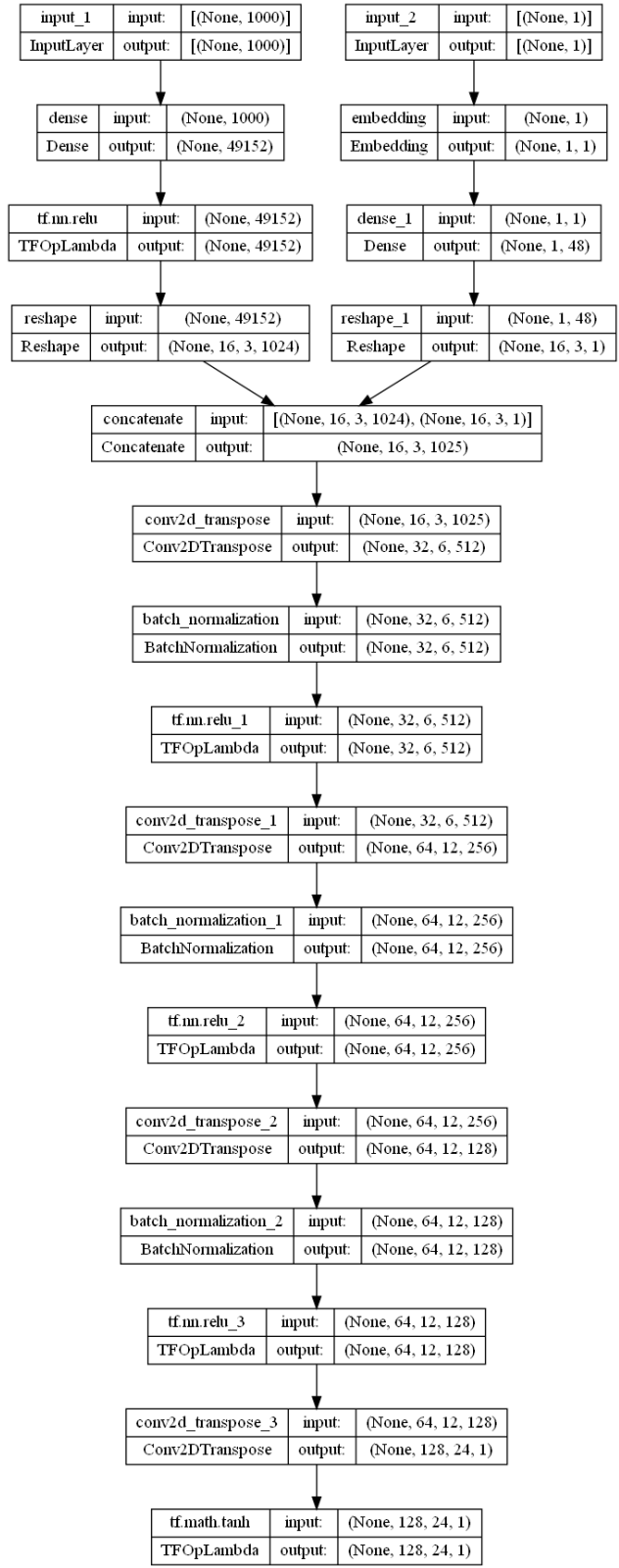

## 10.4 Training and evaluation

After constructing the ACGAN the next step is to train the model on the processed dataset. Figure 10.11 shows a custom designed training loop designed to train the ACGAN model and evaluate its performance simutaniously.

*Figure 10.11: Defining the training loop for the ACGAN*

```python
def train(epochs):
    global x_train

    nb_test = testing_size
    nb_train = training_size

    train_history = defaultdict(list)
    test_history = defaultdict(list)

    for epoch in range(epochs):

        print("Epoch {} of {}".format(epoch + 1, epochs))
        nb_batches = int(nb_train / BATCH_SIZE)
        progress_bar = Progbar(target=nb_batches)

        epoch_gen_loss = []
        epoch_disc_real_loss = []
        epoch_disc_fake_loss = []

        for index in range(nb_batches):
            progress_bar.update(index)

            noise = np.random.normal(0, 0.02, (BATCH_SIZE, LATENT_DIM))

            image_batch = []
            for _ in range(BATCH_SIZE):
                try:
                    image_batch.append(next(x_train))
                except StopIteration:
                    x_train = load_and_preprocess_audio(x_train_file_paths)
                    image_batch.append(next(x_train))
            image_batch = np.array(image_batch)

            label_batch = y_train[index * BATCH_SIZE : (index + 1) * BATCH_SIZE]

            sampled_labels = np.random.randint(0, 1, BATCH_SIZE)

            generated_images = generator.predict(
```

```python
                [sampled_labels.reshape((-1, 1)), noise], verbose=0
        )

        # X = np.concatenate((image_batch, generated_images))
        # y = np.array([1] * BATCH_SIZE + [0] * BATCH_SIZE)

        # aux_y = np.concatenate((label_batch, sampled_labels))

        real_actual_labels = np.array([1] * BATCH_SIZE)
        fake_actual_labels = np.array([0] * BATCH_SIZE)

        epoch_disc_real_loss.append(
            discriminator.train_on_batch(
                image_batch, [real_actual_labels, label_batch]
            )
        )
        epoch_disc_fake_loss.append(
            discriminator.train_on_batch(
                generated_images, [fake_actual_labels, sampled_labels]
            )
        )

        noise = np.random.normal(
            0, 0.02, (NUM_CLASSES * BATCH_SIZE, LATENT_DIM)
        )
        sampled_labels = np.random.randint(
            0, NUM_CLASSES - 1, NUM_CLASSES * BATCH_SIZE
        )

        trick = np.ones(NUM_CLASSES * BATCH_SIZE)

        epoch_gen_loss.append(
            combined.train_on_batch(
                [sampled_labels.reshape((-1, 1)), noise], [trick,
sampled_labels]
            )
        )

    print("\nTesting for epoch {}:".format(epoch + 1))
    noise = np.random.normal(0, 0.02, (nb_test, LATENT_DIM))

    sampled_labels = np.random.randint(0, NUM_CLASSES - 1, nb_test)
    generated_images = generator.predict(
        [sampled_labels.reshape((-1, 1)), noise], verbose=False
    )
```

```python
    # X = np.concatenate((x_test, generated_images))
    # y = np.array([1] * nb_test + [0] * nb_test)
    # aux_y = np.concatenate((y_test, sampled_labels), axis=0)

    real_actual_labels = np.array([1] * nb_test)
    fake_actual_labels = np.array([0] * nb_test)

    discriminator_test_real_loss = discriminator.evaluate(
        x_test, [real_actual_labels, y_test], verbose=False
    )
    discriminator_test_fake_loss = discriminator.evaluate(
        generated_images, [fake_actual_labels, sampled_labels], verbose=False
    )

    discriminator_train_real_loss = np.mean(np.array(epoch_disc_real_loss),
axis=0)
    discriminator_train_fake_loss = np.mean(np.array(epoch_disc_fake_loss),
axis=0)

    noise = np.random.normal(0, 0.02, (2 * nb_test, LATENT_DIM))
    sampled_labels = np.random.randint(0, NUM_CLASSES - 1, NUM_CLASSES *
nb_test)

    trick = np.ones(NUM_CLASSES * nb_test)

    generator_test_loss = combined.evaluate(
        [sampled_labels.reshape((-1, 1)), noise],
        [trick, sampled_labels],
        verbose=False,
    )

    generator_train_loss = np.mean(np.array(epoch_gen_loss), axis=0)

    train_history["generator"].append(generator_train_loss)
    train_history["discriminator_real"].append(discriminator_train_real_loss)
    train_history["discriminator_fake"].append(discriminator_train_fake_loss)

    test_history["generator"].append(generator_test_loss)
    test_history["discriminator_real"].append(discriminator_test_real_loss)
    test_history["discriminator_fake"].append(discriminator_test_fake_loss)

    # Display generator loss
    header = "{0:<27s} | {1:10s} | {2:14s} | {3:10s} ".format(
        "component", "total loss", "real/fake loss", "class loss"
```

```python
    )
    print(header)
    print("-" * len(header))
    ROW_FMT = "{0:<27s} | {1:10.2f} | {2:14.2f} | {3:10.2f} "

    print(ROW_FMT.format("generator (train)", *train_history["generator"][-
1]))
    print(ROW_FMT.format("generator (test)", *test_history["generator"][-1]))
    print("-" * len(header))

    # Display discriminator loss
    header = "{0:<27s} | {1:10s} | {2:14s} | {3:10s} | {4:18s} |
{5:14s}".format(
        "component",
        "total loss",
        "real/fake loss",
        "class loss",
        "real/fake accuracy",
        "class accuracy",
    )

    print(header)
    print("-" * len(header))
    ROW_FMT = "{0:<27s} | {1:10.2f} | {2:14.2f} | {3:10.2f} | {4:18.2f} |
{5:14.2f}"

    print(
        ROW_FMT.format(
            "discriminator real (train)",
*train_history["discriminator_real"][-1]
        )
    )
    print(
        ROW_FMT.format(
            "discriminator fake (train)",
*train_history["discriminator_fake"][-1]
        )
    )
    print(
        ROW_FMT.format(
            "discriminator real (test)",
*test_history["discriminator_real"][-1]
        )
    )
    print(
```

```python
            ROW_FMT.format(
                "discriminator fake (test)",
*test_history["discriminator_fake"][-1]
            )
        )
        print("-" * len(header))

        # Save the model every 10 epochs
        if (epoch + 1) % 10 == 0:
            checkpoint.save(file_prefix=checkpoint_prefix)
            pickle.dump(
                {"train": train_history, "test": test_history},
                open(
                    os.path.join(training_checkpoints_directory, "acgan-
history.pkl"),
                    "wb",
                ),
            )

        r, c = 2, 2
        noise = np.random.normal(0, 0.02, (r * c, LATENT_DIM))

        sampled_labels = np.array([num for _ in range(r) for num in range(c)])
        sampled_labels = sampled_labels.reshape((-1, 1))
        gen_imgs = generator.predict([sampled_labels, noise])
        gen_imgs = 0.5 * gen_imgs + 0.5

        fig, axs = plt.subplots(r, c)
        cnt = 0
        for i in range(r):
            for j in range(c):
                # Save Audio
                mel_inv = feature_to_audio(gen_imgs[cnt, :, :, :])
                sf.write(
                    os.path.join(generated_audio_path,
f"epoch_{epoch}_{cnt}.wav"),
                    mel_inv,
                    fs,
                )
                # Add image to figure
                axs[i, j].imshow(gen_imgs[cnt, :, :, 0], cmap="gray")
                axs[i, j].axis("off")
                cnt += 1
        fig.savefig(os.path.join(generated_images_path, "%d.png" % epoch))
        plt.close()
```

```
    pickle.dump(
        {"train": train_history, "test": test_history},
        open(os.path.join(training_checkpoints_directory, "acgan-history.pkl"),
"wb"),
    )
```

### 10.4.1 Training

For each batch, the image batch and respective labels of those spectrograms are generated then the generator model is tasked with generating spectrograms for different class labels. The discriminator is training on the real images and their labels and its loss during that training is labeled as real loss representing the class and validation loss on real data. Then the model is trained on the generated samples where the loss measured is labeled fake loss. Both of the losses are recorded to be displayed after each epoch.

*Figure 10.12: Training the discriminator and recording the real and fake losses*

```
        for index in range(nb_batches):
            progress_bar.update(index)

            noise = np.random.normal(0, 0.02, (BATCH_SIZE, LATENT_DIM))

            image_batch = []
            for _ in range(BATCH_SIZE):
                try:
                    image_batch.append(next(x_train))
                except StopIteration:
                    x_train = load_and_preprocess_audio(x_train_file_paths)
                    image_batch.append(next(x_train))
            image_batch = np.array(image_batch)

            label_batch = y_train[index * BATCH_SIZE : (index + 1) * BATCH_SIZE]

            sampled_labels = np.random.randint(0, 1, BATCH_SIZE)

            generated_images = generator.predict(
                [sampled_labels.reshape((-1, 1)), noise], verbose=0
            )

            # X = np.concatenate((image_batch, generated_images))
            # y = np.array([1] * BATCH_SIZE + [0] * BATCH_SIZE)

            # aux_y = np.concatenate((label_batch, sampled_labels))
```

```
        real_actual_labels = np.array([1] * BATCH_SIZE)
        fake_actual_labels = np.array([0] * BATCH_SIZE)

        epoch_disc_real_loss.append(
            discriminator.train_on_batch(
                image_batch, [real_actual_labels, label_batch]
            )
        )
        epoch_disc_fake_loss.append(
            discriminator.train_on_batch(
                generated_images, [fake_actual_labels, sampled_labels]
            )
        )
```

Then after the discriminator's weight has been updated with backpropagation the composite model is trained with generated noise and randomly selected labels. Since the composite model consists of both the generator and discriminator the noise and class labels will be converted into generated images that the discriminator will then label. Since the two models are connected the loss of the discriminator will backpropagate to the generator without altering the weights of the discriminator. Since the discriminator was updated before the generator the loss will improve the generator's performance.

*Figure 10.13: Training the generator*

```
        noise = np.random.normal(
            0, 0.02, (NUM_CLASSES * BATCH_SIZE, LATENT_DIM)
        )
        sampled_labels = np.random.randint(
            0, NUM_CLASSES - 1, NUM_CLASSES * BATCH_SIZE
        )

        trick = np.ones(NUM_CLASSES * BATCH_SIZE)

        epoch_gen_loss.append(
            combined.train_on_batch(
                [sampled_labels.reshape((-1, 1)), noise], [trick,
sampled_labels]
            )
        )
```

## 10.4.2 Validation

At the end of each epoch, both the generator and discriminator are tested using the testing data that neither has seen before. The testing procedure is identical to the training procedure with the exception that the weights are not updated.

*Figure 10.14: Validating the performance of the generator and discriminator*

```python
        print("\nTesting for epoch {}:".format(epoch + 1))
        noise = np.random.normal(0, 0.02, (nb_test, LATENT_DIM))

        sampled_labels = np.random.randint(0, NUM_CLASSES - 1, nb_test)
        generated_images = generator.predict(
            [sampled_labels.reshape((-1, 1)), noise], verbose=False
        )

        # X = np.concatenate((x_test, generated_images))
        # y = np.array([1] * nb_test + [0] * nb_test)
        # aux_y = np.concatenate((y_test, sampled_labels), axis=0)

        real_actual_labels = np.array([1] * nb_test)
        fake_actual_labels = np.array([0] * nb_test)

        discriminator_test_real_loss = discriminator.evaluate(
            x_test, [real_actual_labels, y_test], verbose=False
        )
        discriminator_test_fake_loss = discriminator.evaluate(
            generated_images, [fake_actual_labels, sampled_labels], verbose=False
        )

        discriminator_train_real_loss = np.mean(np.array(epoch_disc_real_loss),
axis=0)
        discriminator_train_fake_loss = np.mean(np.array(epoch_disc_fake_loss),
axis=0)

        noise = np.random.normal(0, 0.02, (2 * nb_test, LATENT_DIM))
        sampled_labels = np.random.randint(0, NUM_CLASSES - 1, NUM_CLASSES *
nb_test)

        trick = np.ones(NUM_CLASSES * nb_test)

        generator_test_loss = combined.evaluate(
            [sampled_labels.reshape((-1, 1)), noise],
            [trick, sampled_labels],
            verbose=False,
        )
```

```python
        generator_train_loss = np.mean(np.array(epoch_gen_loss), axis=0)

        train_history["generator"].append(generator_train_loss)
        train_history["discriminator_real"].append(discriminator_train_real_loss)
        train_history["discriminator_fake"].append(discriminator_train_fake_loss)

        test_history["generator"].append(generator_test_loss)
        test_history["discriminator_real"].append(discriminator_test_real_loss)
        test_history["discriminator_fake"].append(discriminator_test_fake_loss)
```

### 10.4.3  Evaluation

Finally, the performance of the ACGAN is evaluated incrementally and displayed to ensure that the training is going smoothly. Furthermore, generated instances are saved and converted to audio to evaluate the quality of the generated audio in real-time.

*Figure 10.15: Displaying the overall loss of the ACGAN and saving generated instances*

```python
        # Display generator loss
        header = "{0:<27s} | {1:10s} | {2:14s} | {3:10s} ".format(
            "component", "total loss", "real/fake loss", "class loss"
        )
        print(header)
        print("-" * len(header))
        ROW_FMT = "{0:<27s} | {1:10.2f} | {2:14.2f} | {3:10.2f} "

        print(ROW_FMT.format("generator (train)", *train_history["generator"][-
1]))
        print(ROW_FMT.format("generator (test)", *test_history["generator"][-1]))
        print("-" * len(header))

        # Display discriminator loss
        header = "{0:<27s} | {1:10s} | {2:14s} | {3:10s} | {4:18s} |
{5:14s}".format(
            "component",
            "total loss",
            "real/fake loss",
            "class loss",
            "real/fake accuracy",
            "class accuracy",
        )

        print(header)
        print("-" * len(header))
```

```python
        ROW_FMT = "{0:<27s} | {1:10.2f} | {2:14.2f} | {3:10.2f} | {4:18.2f} | {5:14.2f}"

        print(
            ROW_FMT.format(
                "discriminator real (train)",
                *train_history["discriminator_real"][-1]
            )
        )
        print(
            ROW_FMT.format(
                "discriminator fake (train)",
                *train_history["discriminator_fake"][-1]
            )
        )
        print(
            ROW_FMT.format(
                "discriminator real (test)",
                *test_history["discriminator_real"][-1]
            )
        )
        print(
            ROW_FMT.format(
                "discriminator fake (test)",
                *test_history["discriminator_fake"][-1]
            )
        )
        print("-" * len(header))

        # Save the model every 10 epochs
        if (epoch + 1) % 10 == 0:
            checkpoint.save(file_prefix=checkpoint_prefix)
            pickle.dump(
                {"train": train_history, "test": test_history},
                open(
                    os.path.join(training_checkpoints_directory, "acgan-history.pkl"),
                    "wb",
                ),
            )

        r, c = 2, 2
        noise = np.random.normal(0, 0.02, (r * c, LATENT_DIM))

        sampled_labels = np.array([num for _ in range(r) for num in range(c)])
```

```python
        sampled_labels = sampled_labels.reshape((-1, 1))
        gen_imgs = generator.predict([sampled_labels, noise])
        gen_imgs = 0.5 * gen_imgs + 0.5

        fig, axs = plt.subplots(r, c)
        cnt = 0
        for i in range(r):
            for j in range(c):
                # Save Audio
                mel_inv = feature_to_audio(gen_imgs[cnt, :, :, :])
                sf.write(
                    os.path.join(generated_audio_path,
f"epoch_{epoch}_{cnt}.wav"),
                    mel_inv,
                    fs,
                )
                # Add image to figure
                axs[i, j].imshow(gen_imgs[cnt, :, :, 0], cmap="gray")
                axs[i, j].axis("off")
                cnt += 1
        fig.savefig(os.path.join(generated_images_path, "%d.png" % epoch))
        plt.close()

    pickle.dump(
        {"train": train_history, "test": test_history},
        open(os.path.join(training_checkpoints_directory, "acgan-history.pkl"),
"wb"),
    )
```

## 11  System Implementation

The end goal of this project is to empower covid-19 detection application that leverages patterns in coughs to detect covid-19. Thus a platform was developed with an embedded CNN trained to detect covid-19 through coughs as shown in figure 11.1.

*Figure 11.1: Home of the covid-19 detection through coughs platform*

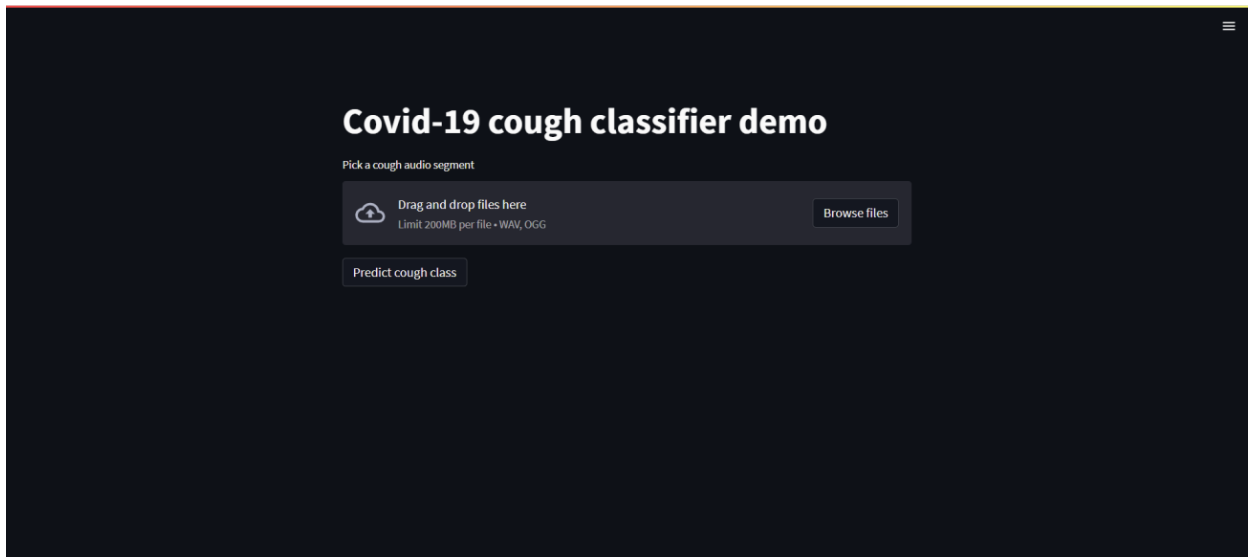

### 11.1  Uploading audio segments

Within the platform, a file upload widget exists that allows the users to upload multiple audio files for classification as shown in figure 11.2. Upon clicking on upload files the user is promoted to select as many audio files as they want as shown in figure 11.3.

*Figure 11.2: Uplaod files widget set to accept audio files only*

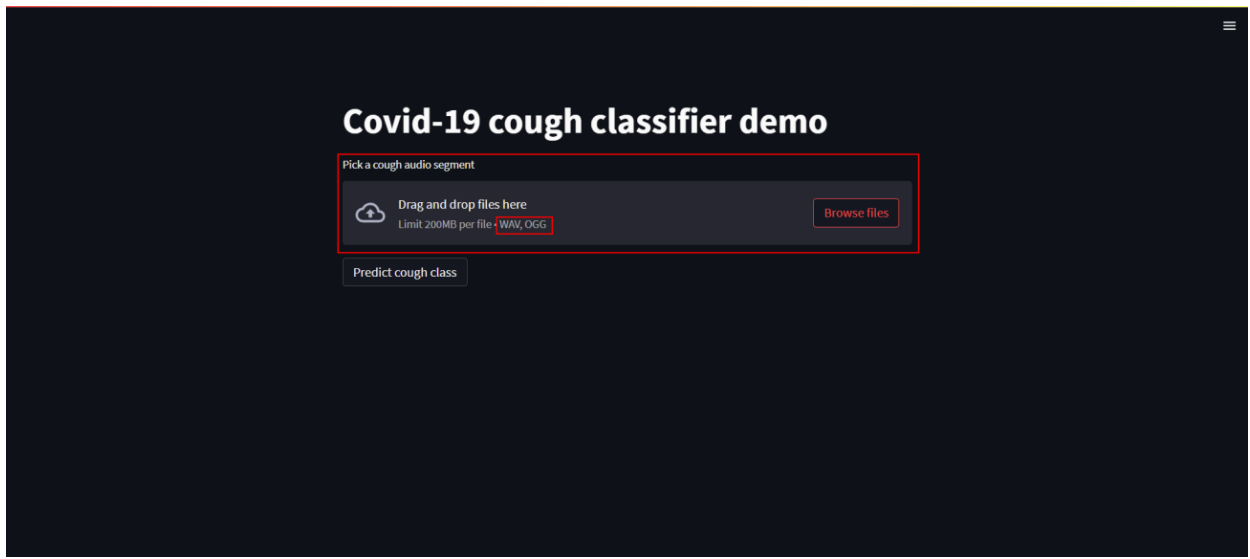

*Figure 11.3: Selecting audio files for classification*

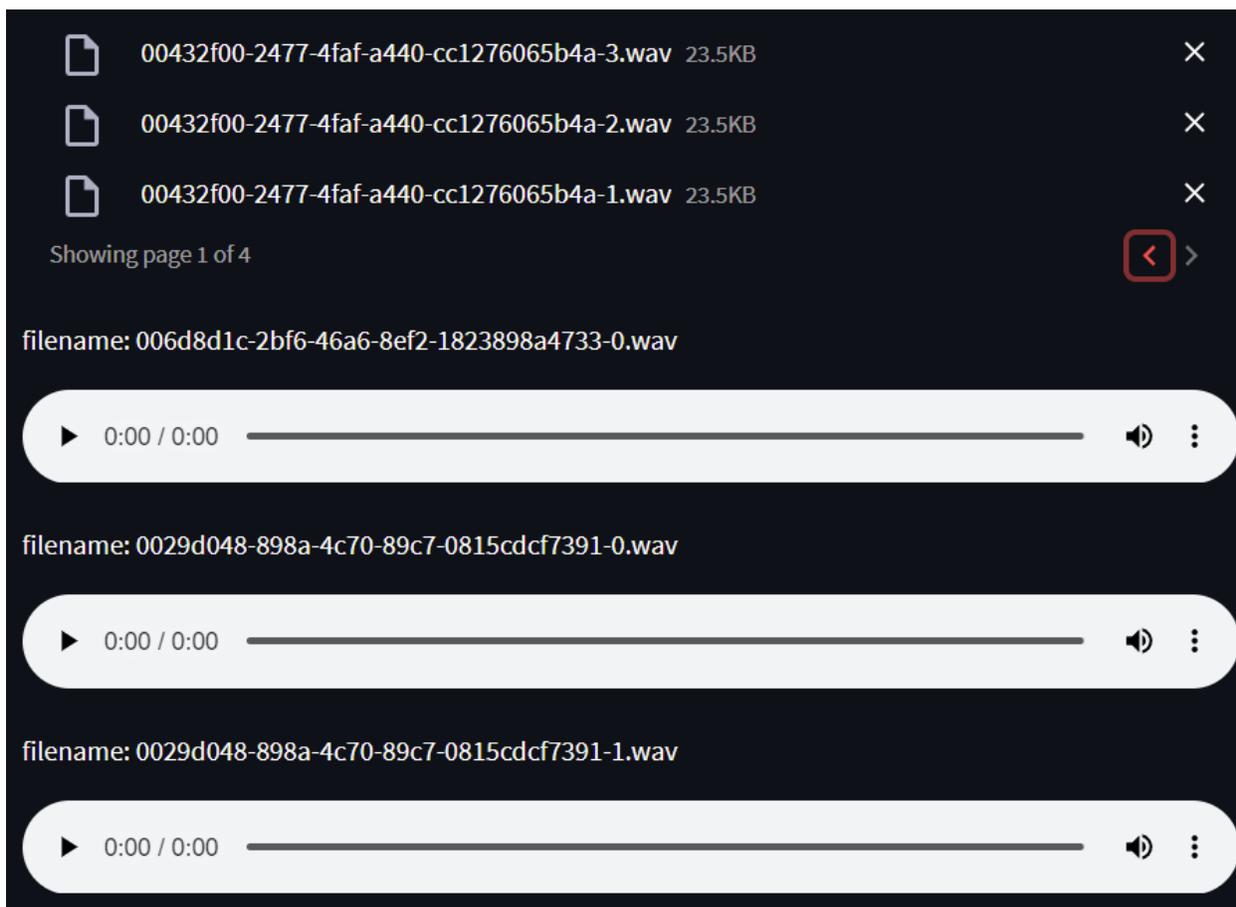

| Name | # | Title | Contributing artists | Album |
|---|---|---|---|---|
| 006d8d1c-2bf6-46a6-8ef2-1823898a4733-0.wav | | | | |
| 0029d048-898a-4c70-89c7-0815cdcf7391-0.wav | | | | |
| 0029d048-898a-4c70-89c7-0815cdcf7391-1.wav | | | | |
| 0029d048-898a-4c70-89c7-0815cdcf7391-2.wav | | | | |
| 0029d048-898a-4c70-89c7-0815cdcf7391-3.wav | | | | |
| 0029d048-898a-4c70-89c7-0815cdcf7391-4.wav | | | | |
| 00432f00-2477-4faf-a440-cc1276065b4a-0.wav | | | | |
| 00432f00-2477-4faf-a440-cc1276065b4a-1.wav | | | | |
| 00432f00-2477-4faf-a440-cc1276065b4a-2.wav | | | | |
| 00432f00-2477-4faf-a440-cc1276065b4a-3.wav | | | | |

## 11.2 Viewing uploaded audio

After uploading the desired audio segment the user can see the file name and play the audio file in real-time before prediction as shown in figure 11.4.

*Figure 11.4: Viewing and playing the uploaded audio*

00432f00-2477-4faf-a440-cc1276065b4a-3.wav 23.5KB ✕

00432f00-2477-4faf-a440-cc1276065b4a-2.wav 23.5KB ✕

00432f00-2477-4faf-a440-cc1276065b4a-1.wav 23.5KB ✕

Showing page 1 of 4 ⟨ ⟩

filename: 006d8d1c-2bf6-46a6-8ef2-1823898a4733-0.wav

▶ 0:00 / 0:00 🔊 ⋮

filename: 0029d048-898a-4c70-89c7-0815cdcf7391-0.wav

▶ 0:00 / 0:00 🔊 ⋮

filename: 0029d048-898a-4c70-89c7-0815cdcf7391-1.wav

▶ 0:00 / 0:00 🔊 ⋮

## 11.3 Classifying audio samples

Scrolling to the end of the audio samples previews the user can find a button named predict cough class, which will run the trained model on the backend to classify the uploaded audio samples and display the results.

*Figure 11.5: Predict cough class button*

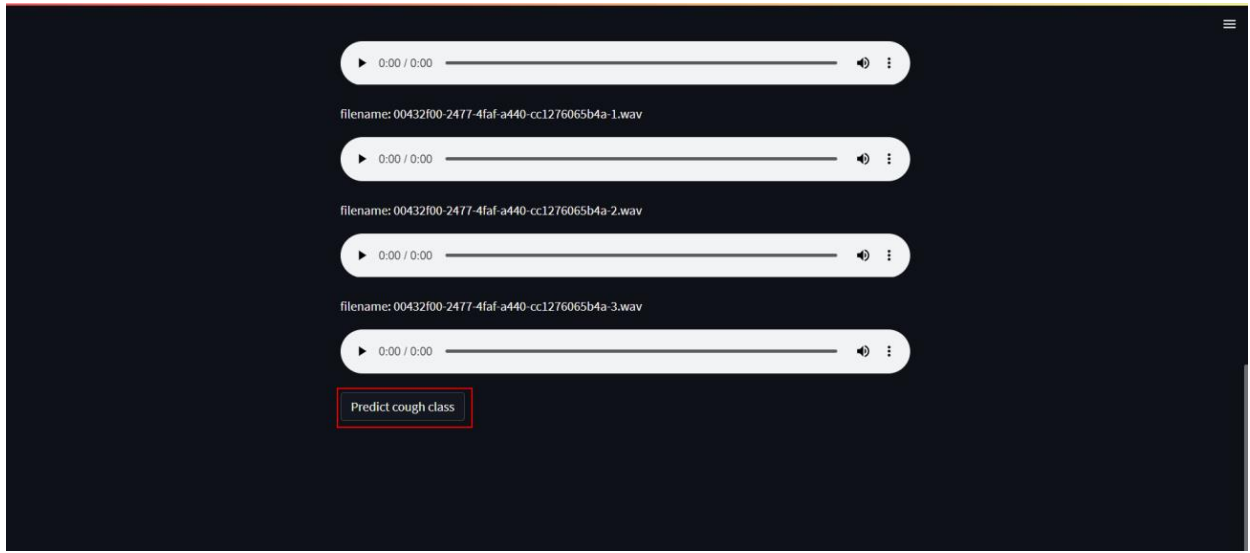

Upon clicking the predict cough button one of two behaviors will happen. If no audio samples were uploaded an error message will be displayed as shown in figure 11.6, otherwise, the model will be executed and a loading spinner will be displayed while the model predicts the classes as shown in figure 11.7.

*Figure 11.6: error message showing that the user did not upload audio segments*

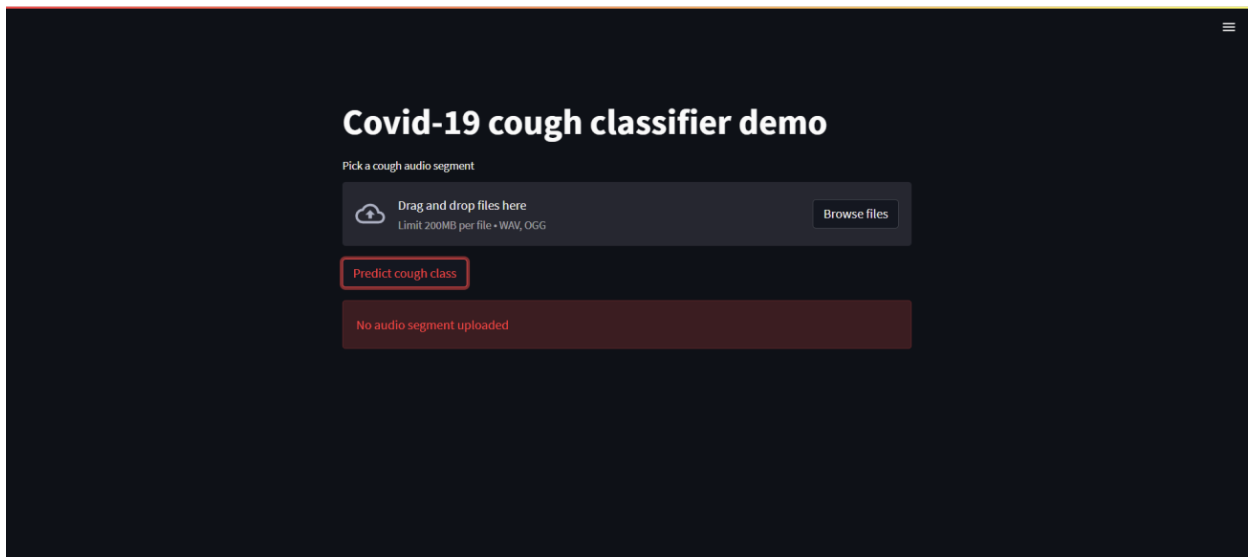

*Figure 11.7: Loading spinner indicating that the model is predicting the output*

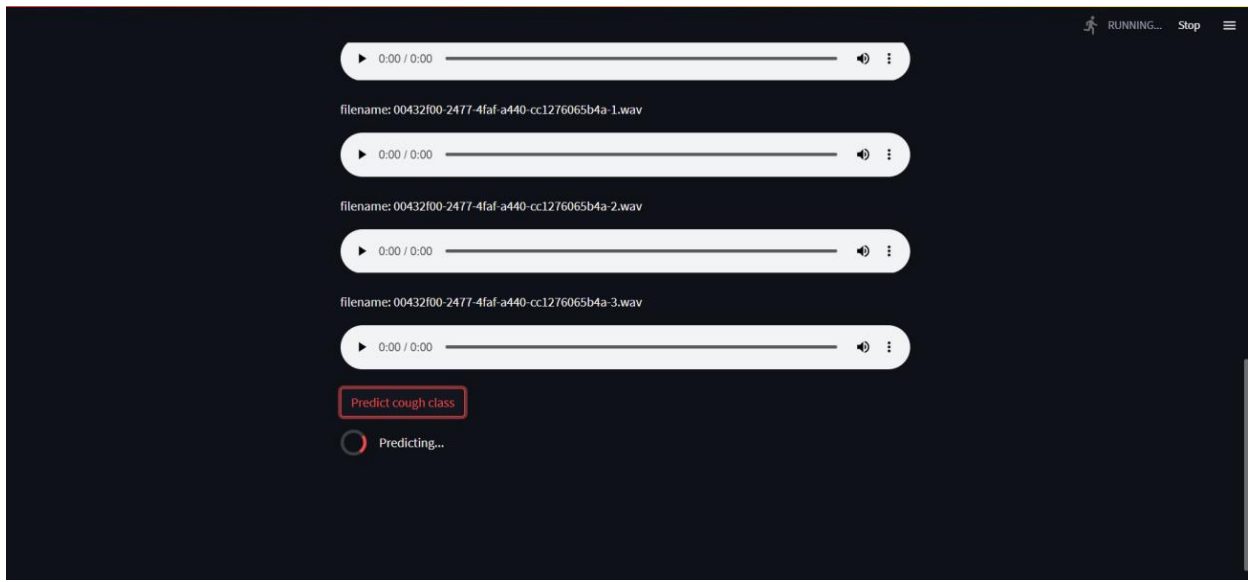

## 11.4  Viewing the predictions of the model

After the model is done predicting the prediction results will be displayed for each uploaded audio sample. As shown in figure 11.8, the probability of the audio belonging to one of the three classes is displayed and the most likely option is highlighted.

*Figure 11.8: Displaying the model predictions*

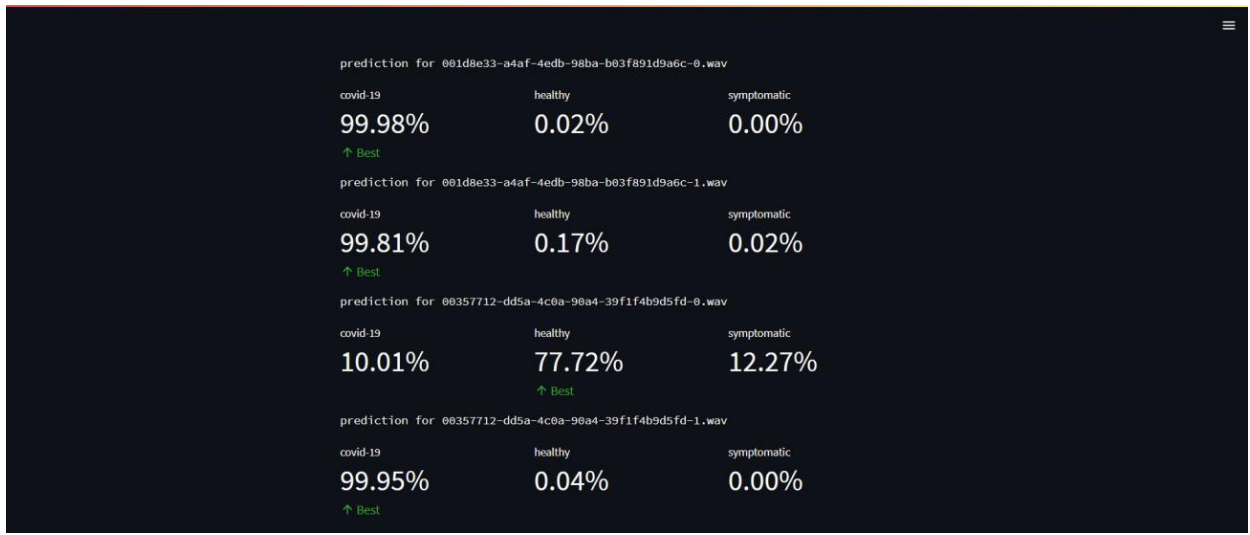

## 12  Conclusion

The spread of the Coronavirus disease 2019 (Covid-19) in the last 2 years has caused a catastrophe in healthcare. Faced with another underrepresented domain in health care, many efforts have turned to the use of GANs (Goodfellow et al., 2014) as a method for mitigating the scarcity within the Covid-19 domain. This paper has reviewed all of the existing GAN efforts in aiding with the Covid-19 detection. The discussed literature used GANs for different tasks that compounded many of the applications of GANs on image data such as image synthesis, semantic segmentation, classification, and saliency map generation. Image synthesis was used for dataset expansion and data anonymization (Jiang et al., 2020; Waheed et al., 2020; Zunair & Hamza, 2021a; Liu et al., 2020). The prevalent approach was image-to-image translation where the GAN is taught to translate a given image from the Covid-19 positive domain to the negative domain. Semantic segmentation was used for lung and lesion region segmentation to improve classifiers' performance (Motamed et al., 2021; Mahapatra & Singh, 2021; Menon et al., 2021; Liu et al., 2020). Most publications used a U-net base to develop the segmentation model with only (Menon et al., 2021) using cycle consistency. Semantic segmentation showed a constant ability to improve image synthesis tasks as demonstrated by (Motamed et al., 2021). Likewise, (Liu et al., 2020) showed that augmenting a dataset for the segmentation model with synthesized data demonstrating the spatial probability map distribution of the target region improves the segmentation maps. Therefore, future applications in either task should consider applying the other task for improved results.

Classification GANs were used to directly classify the label of the given image (Waheed et al., 2020; Motamed et al., 2021). While this application received less attention it demonstrated the application of GAN in a semi-supervised setting which is a robust approach to scarce datasets. Saliency map generation provided an insight into the interworkings of the model and can be used as a sanity check. (Dravid et al ., 2021) demonstrated that GANs can be leveraged to generate a rich activation map highlighting the features unique to a given class. There is great promise in this approach for both new and well-known domains. The activation map can be used to ensure that a model is focusing on the correct feature and in new domains, this representation can provide insights on the domain as demonstrated by (Zunair & Hamza, 2021b). The proposed GAN for saliency map generation can be further improved to generate saliency maps in both directions, source to target and target to source. The two generated maps can then be compared for a better feature representation.

Existing literature on time series prediction for anticipating the Covid-19 spread has been reviewed where GANs were used as a computationally efficient alternative to Agent-based and SEIRS-type models, modeling the Spatio-temporal space of the environment and predicting the spread of the disease (Silva et al., 2021a; Silva et al., 2021b). While this approach was explored in computational physics, it is new to virus modeling and can offer a better method for anticipating the impact of a virus and helping mitigate the spread.

All of the reviewed papers used radiological data namely Chest X-rays (CXR) and Computed Tomography (CT) as the target domain for the model. The paper distribution by selected data, shown in Figure 2, shows that the publications pivoted towards using CXR data over CT data which could be due to the availability of CXR images in general; however, the adoption of both radiological data was the norm. Overall, radiological data was predominantly adopted as it offered reliable feature representation of Covid-19 symptoms over reverse-transcription polymerase chain reaction (RT-PCR) (Ai et el., 2020; Fang et al., 2020; Bernheim et al., 2020; Chung et al., 2020; Zhao et al., 2020b; Kanne et al., 2020; Guan et al., 2020; Ai et al., 2020; Huang et al., 2020). However, RT-PCR and radiological imaging are not the only reliable methods for Covid-19 detection. Many publications, such as (Darici et al., 2022; Sharma et al., 2021; Muguli et al., 2021; Orlandic et al., 2021; Södergren et al., 2021), have shown promising results in detecting a Covid-19 patient using the person's cough. (Darici et al., 2022) tested multiple models for the task of detecting covid given cough audio. The study found a possibility for bias given the limited dataset and concluded that more data is required to improve the performance of the model. (Lella & PJA, 2021) have collected around five thousand audio samples for the research with only 300 of those samples declared as Covid-19 positive patients. Despite the clear data imbalances promising publications and challenges were released. The DiCOVA challenge, its first launch (Muguli et al., 2021) and second (Sharma et al., 2021), was designed to encourage researchers to analyze recordings of people infected and not infected with covid. However, much like the previous datasets, the ones featured in the tracks were fairly biased with the first track dataset having 965 samples only 172 of which are Covid-19 positive, and the second track having 471 only 71 of which are covid positive (Sharma et al., 2021). And indeed, this bias was reflected in the performance of teams in the challenge as they stated in their paper (Sharma et al., 2022) that the top 4 teams fell short of the $\geq 70\%$ (at a specificity of 95%) stated by the World Health Organization. Although a simple combination of the scores from models developed by those teams meets the stated goal the authors state "there is scope for further development in future" (Sharma et al., 2021). But despite the potential of this new approach to Covid-19 detection and data imbalance, to the best of our knowledge, there has not been any published application of GANs in this specific sub-domain.

Focusing on the discussed literature, no consistent evaluation matrix was used by different authors. Instead, each paper would apply the proposed GAN in the targeted use case to demonstrate its effectiveness. And while the object of each paper within a discussed section was identical the proposed use case differed which in turn led to different evaluation methods. For example, (Motamed et al., 2021; Mahapatra & Singh, 2021; Menon et al., 2021; Liu et al., 2020) had the same objective: developing a segmentation model for lung segmentation within the Covid-19 domain; however, each paper evaluated their model different. (Motamed et al., 2021) evaluated the accuracy of the model on 50 excluded manually segmented CXR using Sørensen–Dice coefficient; (Mahapatra & Singh, 2021) evaluated the model against other segmentation models; (Menon et al., 2021) evaluated the model by using it to augment the dataset of selected classifiers; and (Liu et al., 2020) proposed an evaluation matrix for evaluating the model's lesion inclusion rate. This disparity in evaluation increases the difficulty of evaluating the proposed models relative to other published work and in turn increases the difficulty of determining the best approach for deployment and future improvement. Likewise, the discussed image synthesis papers collectively wanted to improve the performance of the classifier by augmenting the given dataset with synthesized data, yet the evaluation methods differed. (Jiang et al., 2020) evaluated the image quality against other models using FID, PSNR, SSIM, RMSE. Evaluated the generated data impact by incrementally adding synthesized data to a U-net and measuring the performance, and evaluated the generated data impact by incrementally adding synthesized data to a U-net and measuring the performance; (Waheed et al., 2020) trained a VGG-16 classifier with and without the synthesized data and compared the performance; and (Zunair & Hamza, 2021a) trained  VGG-16, ResNet-50, DenseNet-102, and (DenseNet121 + BGT)  with and without the synthesized data and compared the performance. While each paper evaluated its model by training a classifier on the augmented dataset, the different choices of classifier architecture and model configuration maintain the shortcoming caused by a lack of a consistent evaluation matrix. To that end, there is a need for an algorithmically based evaluation matrix that objectively evaluates the GAN designed for each specific task.

This work explored the application of GANs in aiding with the Covid-19 detection through coughs. The existing literature showed promising results in deploying gans to synthesize chest x-rays and CT scans. Furthermore, the existing classifiers build to detect Covid-19 showed promising results but reported data imbalance that resulted in classification bias. To mitigate such bias, this work developed a generative adversarial network to augment and synthesize covid-19 coughs to expand the existing datasets and enable classifier better detect covid-19.